\documentclass[]{aa}
\usepackage{graphics}
\begin{document}
\thesaurus{3 (11.02.2, 13.07.2)}

\setlength{\tabcolsep}{0.1cm}
\setcounter{totalnumber}{12}
\renewcommand{\topfraction}{0.98}
\renewcommand{\bottomfraction}{0.98}
\renewcommand{\floatpagefraction}{0.98}
\renewcommand{\figurename}{\small Fig.}
\renewcommand{\tablename}{\small Table}
\setlength{\floatsep}{0cm}
\setlength{\intextsep}{0cm}
\title{The Temporal Characteristics of the TeV
Gamma-Radiation from Mkn~501 in 1997, I: Data from the Stereoscopic
Imaging Atmospheric Cherenkov Telescope System of HEGRA}
\author{F. Aharonian\inst1,
A.G. Akhperjanian\inst{7},
J.A.~Barrio\inst{2,3},
K.~Bernl\"ohr\inst{1,8},
H. Bojahr\inst{6},
J.L. Contreras\inst{3},
J. Cortina\inst{3},
A. Daum\inst{1},
T. Deckers\inst{5},
V. Fonseca\inst{3},
J.C. Gonzalez\inst{3},
G. Heinzelmann\inst{4},
M. Hemberger\inst{1},
G. Hermann\inst{1},
M. Hess\inst{1},
A. Heusler\inst{1},
W. Hofmann\inst{1},
H. Hohl\inst{6},
D. Horns\inst{4},
A. Ibarra\inst{3},
R. Kankanyan\inst{1},
O. Kirstein\inst{5},
C. K\"ohler\inst{1},
A. Konopelko\inst{1},
H. Kornmeyer\inst{2},
D. Kranich\inst{2},
H. Krawczynski\inst{1,4},
H. Lampeitl\inst{1},
A. Lindner\inst{4},
E. Lorenz\inst{2},
N. Magnussen\inst{6},
H. Meyer\inst{6},
R. Mirzoyan\inst{2},
A. Moralejo\inst{3},
L. Padilla\inst{3},
M. Panter\inst{1},
D. Petry\inst{2,6,9},
R. Plaga\inst{2},
A. Plyasheshnikov\inst{1},
J. Prahl\inst{4},
G. P\"uhlhofer\inst{1},
G. Rauterberg\inst{5},
C. Renault\inst{1},
W. Rhode\inst{6},
V. Sahakian\inst{7},
M. Samorski\inst{5},
D. Schmele\inst{4},
F. Schr\"oder\inst{6},
W. Stamm\inst{5},
H. V\"olk\inst{1},
B. Wiebel-Sooth\inst{6},
C. Wiedner\inst{1},
M. Willmer\inst{5},
H. Wirth\inst{1}}
\institute{Max Planck Institut f\"ur Kernphysik,
Postfach 103980, D-69029 Heidelberg, Germany \and
Max Planck Institut f\"ur Physik, F\"ohringer Ring
6, D-80805 M\"unchen, Germany \and
Universidad Complutense, Facultad de Ciencias
F\'{\i}sicas, Ciudad Universitaria, E-28040 Madrid, Spain \and
Universit\"at Hamburg, II. Institut fuer
Experimentalphysik, Luruper Chausse 149,
D-22761 Hamburg, Germany \and
Universit\"at Kiel, Institut f\"ur Kernphysik,
Olshausenstr. 40, D-24118 Kiel, Germany \and
Universit\"at Wuppertal, Fachbereich Physik,
Gau{\ss}str.20, D-42097 Wuppertal, Germany \and
Yerevan Physics Institute, Alikhanian Br. 2, 375036 Yerevan, Armenia \and
Now at Forschungszentrum Karlsruhe, P.O. Box 3640, D-76021 Karlsruhe \and
Now at Universidad Aut\'{o}noma de Barcelona, 
Institut de F\'{i}sica d'Altes Energies, E-08193 Bellaterra, Spain}
\offprints{Henric Krawczynski, email adress: Henric.Krawczynski@mpi-hd.mpg.de}
\date{Received August 13, 1998 ; accepted ---}
\authorrunning{F. Aharonian et al.}
\titlerunning {TeV Characteristics of Mkn~501}
\maketitle
\begin{abstract}
During 1997, the BL Lac Object Mkn~501 was
the brightest known object in the TeV $\gamma$-ray sky.
The emission was characterized by dramatic variations in
intensity with a mean flux exceeding by a factor of 
three the steady $\gamma$-ray flux of 
the Crab Nebula.
The stereoscopic HEGRA system of 4 Imaging Atmospheric Cherenkov
Telescopes, with an energy threshold of about 500 GeV,
an angular resolution of 0.1$^\circ$, an energy resolution 
of $20 \%$, and a flux sensitivity $\nu\,F_\nu$ at 1\, TeV of
$10^{-11} \, \rm ergs/cm^2 sec$  $\simeq$\,1/4 Crab for 1 hour 
of observation time (S/$\sqrt{\textrm{B}}$=5$\sigma$),
has been used in 1997 for a comprehensive 
study of the spectral and temporal characteristics of 
the TeV $\gamma$-ray emission from Mkn~501 on time scales 
of several hours or less. 
In this paper (Part I) the $\gamma$-ray fluxes 
and spectra on a diurnal basis during the period 
March to October 1997 are presented. 
Furthermore, the
correlation of the TeV emission 
with the flux measured by the 
RXTE All Sky Monitor 
in the energy range from 2 to 12 keV are studied. 
Finally the implications of these results on the physics of 
relativistic jets in BL Lac objects are briefly discussed.
The companion paper (Part II) 
describes the results from the stand alone telescopes
CT1 and CT2.
\end{abstract}

\section{Introduction}
\label{intro}

Mrk~501 was discovered as a source of TeV-$\gamma$-radiation in 1995 by 
the Whipple group (Quinn et al.\ \cite{Quin:96}).
The observation was confirmed later by the HEGRA collaboration 
(Bradbury et al.\ \cite{Brad:97}). 
Together with two other extragalactic 
TeV $\gamma$-ray sources detected so far,   
Mrk~421 (Punch et al.\ \cite{Punc:92}; Petry et al.\ \cite{Petr:96}) and 1ES~2344+514 (Catanese et al.\ \cite{Catan:98}),
Mrk~501 belongs to a sub-population of Active Galactic
Nuclei (AGNs), the so-called BL Lac objects.
Flux variability on various time scales, ranging
from dramatic flares of Mrk~421 in May 1996 with durations 
of about 1 h (Gaidos et al.\ \cite{Gaid:96}) to a state of high flaring
activity of Mrk~501 which lasted several months (e.g. Protheroe et al.\ \cite{Prot:97}), 
is a characteristic feature of the TeV emission 
observed from BL Lac objects.
This agrees well with the general properties of BL Lac objects --  
highly variable AGNs without significant optical line emission, 
but showing a strong nonthermal (synchrotron) 
component of radiation from radio to X-ray wavelengths 
(e.g. Urry \& Padovani \cite{PadUry:95}). 

The correlated flares of BL Lac objects in the keV energy band and 
in the TeV energy band, discovered for the
first time during simultaneous observations of Mrk~421 by the 
Whipple and ASCA detectors (Takahashi et al.\ \cite{ASCA_421:96}; Buckley et al.\  
\cite{Buck:96}),
strongly support the commonly accepted view that both components 
originate in a relativistic jet, with Doppler 
factors $\delta_{\rm j} \geq 5$,  due to synchrotron and 
inverse Compton (IC) radiation of the same population 
of ultrarelativistic electrons 
(for a review see e.g. Ulrich et al.\ \cite{UUM:1997}).
Since in the Thomson regime 
the IC cooling time  
$t_{\rm IC}=-E_{\rm e}/({\rm d}E_{\rm e}/{\rm d}t)$ is proportional to 
$1/E_{\rm e}$, and since the Compton scattering boosts ambient 
photons with energies $\epsilon_{\rm 0}$ up to 
$E_\gamma \propto \epsilon_{\rm 0} \cdot E_{\rm e}^2$, the characteristic 
time of $\gamma$-ray emission
decreases with energy as 
$\propto E_\gamma^{-1/2}$. 
This explains in a natural way the less dramatic variations of the
MeV/GeV $\gamma$-ray fluxes during the keV/TeV flares; 
the relatively low energy 
electrons, responsible for the GeV IC photons as well as for 
the optical/UV synchrotron radiation 
do not respond as rapidly to changes of the physical
conditions in the jets as the high energy electrons do. 
In addition, the expected hard spectra of IC radiation below 100 GeV
explain the low fluxes of GeV $\gamma$-rays from Mrk~421,
and their non-detection by EGRET in the case of Mkn~501 and 1ES~2344+514.      
This implies that the VHE $\gamma$-ray region,  
combined with X-ray observations, is likely to be the most important
window of the electromagnetic spectrum to infer
the highly non-stationary processes of particle acceleration and their 
radiation in BL Lac objects. 
Imaging Atmospheric Cherenkov Telescope (IACT)
detectors, characterized by large 
effective detection areas of $\sim\,10^5$\,m$^2$
and an effective suppression of the background of cosmic rays,
are well suited to access this very informative ``TeV'' channel.
This was convincingly demonstrated by the Mkn~501 observations
with several Cherenkov telescopes located in the Northern
Hemisphere during the extreme activity of the source in 1997 
(Protheroe et al.\ \cite{Prot:97}).

During the first two years after its discovery as a TeV $\gamma$-ray source,
Mkn~501 showed rather low fluxes at a level significantly below 
the Crab flux (Quinn et al.\ \cite{Quin:96}; Bradbury et al. \cite{Brad:97}).
However, in March 1997 the source went into a state of
highly variable and strong emission with maximum fluxes
roughly 10 times that of the Crab.
According to the All Sky Monitor
(ASM) on board the {\it Rossi X-Ray
Timing Explorer} (RXTE) (Remillard \& Levine \linebreak[4] \cite{Remi:97}), 
the high X-ray activity of the source
started in March 1997 and continued until October 1997.
Apparently the period of high activity coincided
with the period of the visibility of the source 
by ground-based optical instruments.
Thus it was possible to continuously monitor the source during this
extremely bright emission period with
several IACTs, i.e.\  with
CAT (Barrau et al.\ \cite{Barr:97}),
HEGRA (Aharonian et al.\ \cite{Ahar:97a}), TACTIC (Bhat et al.\ \cite{Bhat:97}), 
Whipple (Catanese et al.\ \cite{Cata:97}),
and the Telescope Array (Hayashida et al.\ \cite{Haya:98}).

The HEGRA experiment is located on the Roque de los Muchachos on the
Canary Island of La Palma,
(lat.\ 28.8$^\circ$ N, long.\ 17.9$^\circ$ W, 2200 m a.s.l.). 
The HEGRA collaboration operates 6 Cherenkov telescopes.
A system of at present four telescopes (telescopes CT3, CT4, CT5, and CT6) is used 
as a single detector for stereoscopic air shower observations (Daum et al.\ \cite{Daum:97}).
The two telescopes, 
CT1 (Mirzoyan et al.\ \cite{Mirz:94}; Rauterberg et al.\ \cite{Raut:95}) and CT2 
(Konopelko et al.\ \cite{konopelko96}), 
are currently operated each as independent detectors. 
The IACT system is characterized by a high sensitivity and excellent spectroscopic
capabilities. The stand alone 
telescopes CT1 and CT2 have been used to considerably extend
the Mkn~501 time coverage, in particular during moonshine periods, when
the stereoscopic system was not operated.

In this paper (Part I) the results obtained from the IACT system data are presented.
The companion paper (Part II) describes in detail the results from
CT1  and CT2 data.

The basic concept of the IACT array is the 
{\it stereoscopic approach} based on 
simultaneous detection of air showers by $\geq 2$ 
telescopes under widely differing viewing angles.
With the stereoscopic technique an angular resolution of 0.1$^\circ$
per photon, an energy resolution of $20\%$ per photon, 
and a suppression of the isotropic cosmic ray background 
on the trigger level and by image analysis
by a factor of the order of 100 is achieved.
Thus $\gamma$-ray observations with unprecedented signal 
to noise ratio and excellent spectroscopic capabilities are possible.
Furthermore, since a hardware trigger requiring
the coincident detection of air showers by at least two telescopes
strongly suppresses triggers caused by the night sky background light or
by local muons, the energy threshold of a stereoscopic telescopes system
is mainly limited by Cherenkov photon statistics.
As a consequence, the IACT system achieves 
an energy threshold as low as 500~GeV despite the relatively small size 
of the telescope mirrors of $8.5 \, \rm m^2$
(the energy threshold is defined as the energy at which the
$\gamma$-ray detection rate peaks for Crab type spectra
with differential photon indices of $\sim -2.5$).
The flux sensitivity of the IACT system for 
episodic TeV $\gamma$-ray phenomena with durations of the order
of 1~h is about 1/4~Crab\footnote{In the case of the Crab Nebula, 
the IACT system  currently gives, 
after cuts, $\simeq$50 $\gamma$-rays per hour over a 
flat background of 6 events per hour.} (for S/$\sqrt{\textrm{B}}$=5$\sigma$), 
which corresponds to a $\nu\,F_\nu$-flux at 1~TeV 
of $\sim 10^{-11} \, \rm erg/cm^2 s$.
This energy flux sensitivity combines nicely with the 
comparable energy flux sensitivities of the current X-ray instruments 
like ASCA, BeppoSAX and RXTE for the study of the high energy emission
of BL Lacs, especially of Mkn~501 and of Mkn~421.
These two sources proved to release a comparable
amount of nonthermal energy in X-rays and in TeV $\gamma$-rays,
with an average energy flux in both channels exceeding 
$\sim 10^{-11} \, \rm erg/cm^2 s$.
During strong flares of these sources with fluxes up to 10~Crab, 
a 2 minute exposure is sufficient for the 
IACT system to detect a statistically 
significant $\gamma$-ray signal, and a 1~h exposure suffices
for a measurement of the differential energy spectrum.

This paper is organized as follows:
Hardware issues are briefly summarized in Section \ref{Hard}.
Analysis methods are presented in Section
\ref{System}.
Subsequently, the results concerning the 
TeV-$\gamma$-ray emission from Mkn~501 in 1997 are discussed.
In Section \ref{lightcurve} the 1997 light curve
of Mkn~501 is presented and possible correlations between
the flux amplitude and the spectral slope are explored.
The most rapid time scales of flux variability
are discussed in Section \ref{variability}.
The correlation of the TeV-fluxes 
with the keV-fluxes as measured with the RXTE All Sky Monitor are studied
in Section \ref{correlation}.
Implications on models of the non-thermal 
$\gamma-$radiation from BL Lac objects
are discussed in Section \ref{Discussion}.
\section{The IACT system of HEGRA}
\label{Hard}
The HEGRA telescope system consists presently of 4, in the near future of 5,
identical IACTs -- one at the center and 3 (in future 4) 
at the corners of a 100 m by 100 m
square area.
The multi-mirror reflector of each telescope has an area of 8.5
m$^2$.
Thirty front aluminized and quartz coated spherical
mirrors of 60\,cm diameter and of
4.9\,m focal length 
are independently mounted on an almost spherical frame
of an alt-azimuth mount, following the design of Davies and Cotton.
Each telescope is equipped with a 271 channel camera of
0.25$^\circ$ pixel size resulting in an effective field of view of 4.3$^\circ$.
The PMT pulses are fed into trigger electronics
and into shapers followed by 120~MHz
flash analog-to-digital converters (FADCs). A multilevel
trigger demands 
{\it at least two adjacent pixels}
in each of {\it at least 2 telescopes} (Bulian et al.\ \cite{Buli:98}).
The topological ``next-neighbor'' condition of two {\it
adjacent} pixels reduces the number of night sky background 
triggers. In the following analysis we use the software trigger condition
of at least two telescopes with images with more than 40 photoelectrons.

At the beginning of each night, the camera is flat-fielded using 
an UV laser at each telescope to illuminate a scintillator via an optical cable. 
The scintillator emits a spectrum with peak emission in the near-UV and blue, similar 
to atmospheric Cherenkov light.
An absolute calibration of the system has been performed with a
direct laser measurement and a calibrated low-power photon detector
(Fra{\ss} et al.\ \cite{Fras:97}).
This measurement has determined the conversion factor from 
photons to FADC counts with an accuracy of 10\%.

The pointing of the telescopes is checked on a regular basis with so
called ``point runs'' (P\"uhlhofer et al.\ \cite{Pueh:97}), where a section of the sky
surrounding a bright star is scanned.
The pointing of each telescope is inferred from the currents
measured in the PMTs surrounding the image of the star. 
After applying the resulting pointing correction function,
an effective pointing accuracy of better than 0.01$^\circ$ is achieved. 
This accuracy has been experimentally confirmed with
$\gamma$-ray data from the Crab Nebula and Mkn~501 (P\"uhlhofer et al.\ \cite{Pueh:97}).

%
%
%
\begin{table}
\caption{\small The IACT system of HEGRA - Hardware changes during 1996 and 1997\label{Hardwaretab}.}
\begin{tabular}{lll}
\hline
&&\\
Date  & Hardware Changes Performed & Data-period\\ 
&&\\
\hline
&&\\
27.11.1996              & Start of 4-telescope system& I\\
& CT3, CT6: 2 pixel trigger,     &\\
& CT4, CT5: 2 pixel topological  &\\
& trigger, single pixel          &\\
& threshold 10 p.e.              &\\ 
&  for all CTs   \\
12.05.1997              & adjustment of telescope mirrors               & II\\   
24.06.1997              & 2 pixel topological trigger and               & III\\
& single pixel threshold 8 p.e.&\\ 
&              for all CTs     &\\
16.10.1997-   & CT4 not operational due to fire & IV\\
15.11.1998&&\\
\hline
\end{tabular}
\end{table}
The first system telescope, 
called CT3, was installed in December 1995.
Subsequently CT4 started operation in July 1996, CT5 in September 1996, 
and CT6 in November 1996. The array of 4 telescopes is 
operational since end of November 1996.
Since then, minor changes of the hardware were carried out.
These are summarized in Table \ref{Hardwaretab}.
In the following, HEGRA data from March 1997 to October 1997 are used.
The relevant hardware changes are 
(i) a mirror adjustment of CT3 and CT4 on May 12th, 1997 and (ii)
the incorporation of the topological next-neighbour trigger
condition on hardware level for CT3 and CT6 as well,
and a reduction of the single pixel trigger threshold
for all IACTs from 10 to 8 photoelectrons on June 24th, 1997.
CT4 and CT5 had been operated with the next-neighbour 
trigger condition from the very beginning.
These changes divide the Mkn~501 1997 data into 3 groups (period I - period III).
In the next subsection it will be shown how the data is corrected for
these changes by the use of detailed Monte Carlo simulations.
\section{Analysis of the IACT system data}
\label{System}
\subsection{Monte Carlo simulations}
\label{MC}
The Monte Carlo simulation (Konopelko et al.\ \cite{Kono:98}) is divided into two
steps. First, the air showers are simulated and the Cherenkov photons
hitting one telescope are stored on
mass storage devices. Thereafter the detector simulation is carried out. 
This method has the advantage that it is
possible to use the same simulated showers with different detector
setups.
%
%
%
\begin{table}
\caption{\small Efficiencies averaged over the wavelength region from 300 to 600\,nm.
\label{MCtab}}
\begin{tabular}{ll}
\hline
&\\
Reason & Efficiency \\
&\\
\hline
&\\
Atmospheric absorption & 0.84 \\
Mirror Reflectivity    & 0.88 \\
Plexiglas and Funnel transmission    & 0.81 \\
Quantum efficiency     & 0.18 \\
Total, Cherenkov Photons to FADC channels& 0.10\\
\hline
\end{tabular}
\end{table}
The air shower simulation is based on the ALTAI code.
The results of the code have been tested against results of the CORSIKA
code which gave for hadron induced air showers 
excellent agreement in all relevant observables, 
e.g., in the predicted detection rates and 
in the distribution of the image parameters (Hemberger \cite{Hemb:98}).
The detector simulation (Hemberger \cite{Hemb:98}) accounts for the absorption
of Che\-ren\-kov
photons in the atmosphere due to ozone absorption, 
Ray\-leigh
scattering and Mie scattering. 
Furthermore, the mirror reflectivity, the mirror point
spread function, and the acceptances of the plexiglass panels and the light
collecting funnels in front of the
cameras are taken into account.
See Table \ref{MCtab} for a summary of the efficiencies which are 
relevant for the simulations. 
In the table only the mean values averaged over the
wavelength region from 300 to 600~nm are given, 
in the simulations the efficiencies depend on the wavelength.
The point spread functions of the telescope mirrors
are extracted from the point runs.
The time-resolved photon to photoelectron conversion by the PMTs is
modeled using a measured PMT pulse shape and a measured single
photoelectron spectrum. 
Finally, the trigger processes and the digitization of the
PMT pulses are simulated in detail.

The simulated events are stored in the same format as the
raw experimental data and
are processed with the same event reconstruction and analysis chain as
the experimental data. Showers induced by photons as well as by 
hydrogen, helium, oxygen, and iron nuclei were
simulated for the zenith angles 
$\theta_\textrm{\small MC}\,=\,$0$^\circ$,$ $20$^\circ$,$ $30$^\circ$, and 45$^\circ$.
For the purpose of comparing the experimental data with the Monte
Carlo predictions, the Monte Carlo events are weighted 
to generate the appropriate spectrum.
The events induced by the proton, helium, oxygen, and
iron nuclei are weighted according to the 
cosmic ray abundances of the corresponding groups
from (e.g.\ Wiebel et al.\ \cite{Wieb:98}). 
For each type of primary particle, and for each zenith angle 
approximately 2\,$\cdot$\,10$^5$ showers have been generated.

The excellent agreement of the observable quantities in the experimental 
data and the Monte Carlo data for cosmic ray-induced showers as well as for
photon-induced showers is described in detail in 
Konopelko et al.\ (\cite{Kono:98}) and in Aharonian et al. (\cite{Ahar:98a}).
The comparisons between data and Monte Carlo which are
relevant for the analysis of the Mkn~501 data of this paper are
discussed in the following 4 subsections.

Three different Monte Carlo event-samples have been generated, 
with different trigger settings and mirror point spread functions,
corresponding to the three data-taking periods.
\subsection{Data sample and data cleaning}
\label{clean}
The analysis described in this paper is based on 110 hours of Mkn~501 data
acquired between March 16th, 1997 and October 1st, 1997 
under optimal weather conditions 
(i.e.\  a clear sky and a humidity less than 90\%), 
with the optimal detector performance, 
and with Mkn~501 being more than 45$^\circ$ above the horizon.
Only data runs where all 4 IACTs where operational and 
in which not more than 20 pixel were defect in any IACT have been admitted
to the analysis. Furthermore the data runs had to satisfy the requirements of
the mean cosmic ray rate deviating by less than 15\% from the
zenith angle dependent expectation value, 
and the width parameter averaged over all
events and all telescopes deviating by less than 6\% 
from the zenith angle dependent expectation value.

The Mkn~501 data were acquired in the so called ``wobble mode'' (Daum et al.\ \cite{Daum:97}).
In this mode the telescopes are pointed into a direction which is shifted by 0.5$^\circ$ 
in declination with respect to the source direction. 
The direction of the shift is reversed for each data run of 
20 minutes duration.
For each run, the solid-angle region located 1$^\circ$ from the 
Mkn~501 location on the opposite side of the camera center is used as OFF region for
estimates concerning the background contamination of the ON region
by cosmic ray-induced showers.
The large angular distance between the ON solid-angle region around
the Mkn~501 direction and the OFF solid-angle region assures a negligible
contamination of the OFF data with Mkn~501 $\gamma$-rays.
The symmetric location of the ON and the OFF region in the camera with
respect to the optical axis and the camera geometry 
assures almost equal background characteristics for both regions.
The zenith angle dependence of the background rate is 
to first order compensated by using as many 
runs with lower-declination OFF regions
as with higher-declination OFF regions.

The overall stability of the detector and the
understanding of the detector performance during the three data periods 
has been tested by comparing several
key Monte Carlo predictions for hadron-induced showers
with the experimental results. 
In the following 3 subsections tests concerning photon-induced showers,
based on $\gamma$-rays from Mkn~501, will be described.

\begin{figure}
\resizebox{\hsize}{!}{\includegraphics{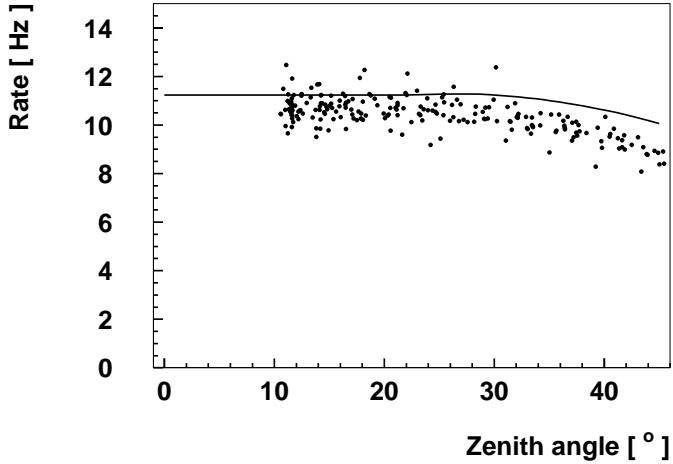}}
\caption{\small
The cosmic ray detection rate of the IACT system (dots) as a function 
of zenith angle (data from data-period I). 
Monte Carlo rate predictions are superimposed (solid line).
The measured and the Monte Carlo data agree with an accuracy of 10\%
(hardware threshold, no cuts).
}
\label{CRfluxfiga} 
\end{figure}

\begin{figure}
\resizebox{\hsize}{!}{\includegraphics{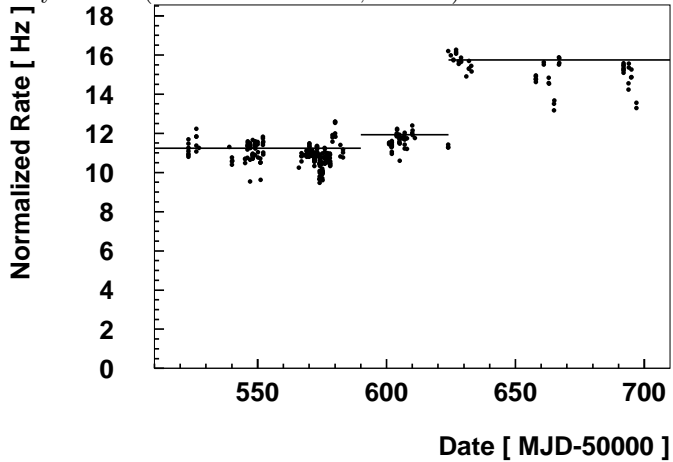}}
\caption{\small
The measured cosmic ray detection rate (dots) and the 
Monte Carlo based predictions (solid line) are shown for all the cosmic ray data of the 
Mkn~501 runs of 1997.
The measured rates have been normalized to a zenith angle of 
15$^\circ$ using an empirical parameterization. The Monte Carlo simulations 
for 0$^\circ$ zenith angle have been used
(hardware threshold, no cuts).}
\label{CRfluxfigb} 
\end{figure}

The observed cosmic ray detection rates have been compared with the
detection rates as inferred from the Monte Carlo simulations
together with the cosmic ray fluxes from the literature.
In Figure \ref{CRfluxfiga} the dependence of the cosmic ray detection
rate on the zenith angle is shown for the first data-period
and the corresponding Monte Carlo data-sample. 
The Monte Carlo describes the 
dependence with an accuracy of 10\%.
In Figure \ref{CRfluxfigb} the measured and predicted rates are shown
for the whole 1997 data-base. The measured rates have been normalized to a
zenith angle of 15$^\circ$ according to the empirical parameterization
shown in Figure \ref{CRfluxfiga}. For all three data periods the Monte Carlo
simulations predict the measured rates with an accuracy of 10\%,
and they accurately describe the relative rate differences between
the data-taking periods. The measured rates within each data-period
show a spread of $4\%$ FWHM, after correcting for the
zenith angle dependence of the rate.
The origin of this small spread is still unclear. The rate deviations do not
correlate with the temperature, the pressure, or the humidity, as  measured at the Nordic
Optical Telescope which is localized within several hundred meters from the
HEGRA site.
Neither is a correlation found with the V-band extinction measured
with the Carlsberg Meridian Circle which is situated at a distance 
of $\sim$500\,m from the HEGRA site.

To summarize, the measurements of the cosmic-ray event rate
prove the stability of the IACT system at a level of
5\% and the event rate is correctly predicted by the Monte Carlo
simulations using the cosmic ray abundances from the literature 
with an accuracy of 10\%.
\subsection{The stereoscopic reconstruction of the direction of
primary particles}
\label{Dir}
%
%
%
\begin{figure*}
\resizebox{12cm}{!}{\includegraphics{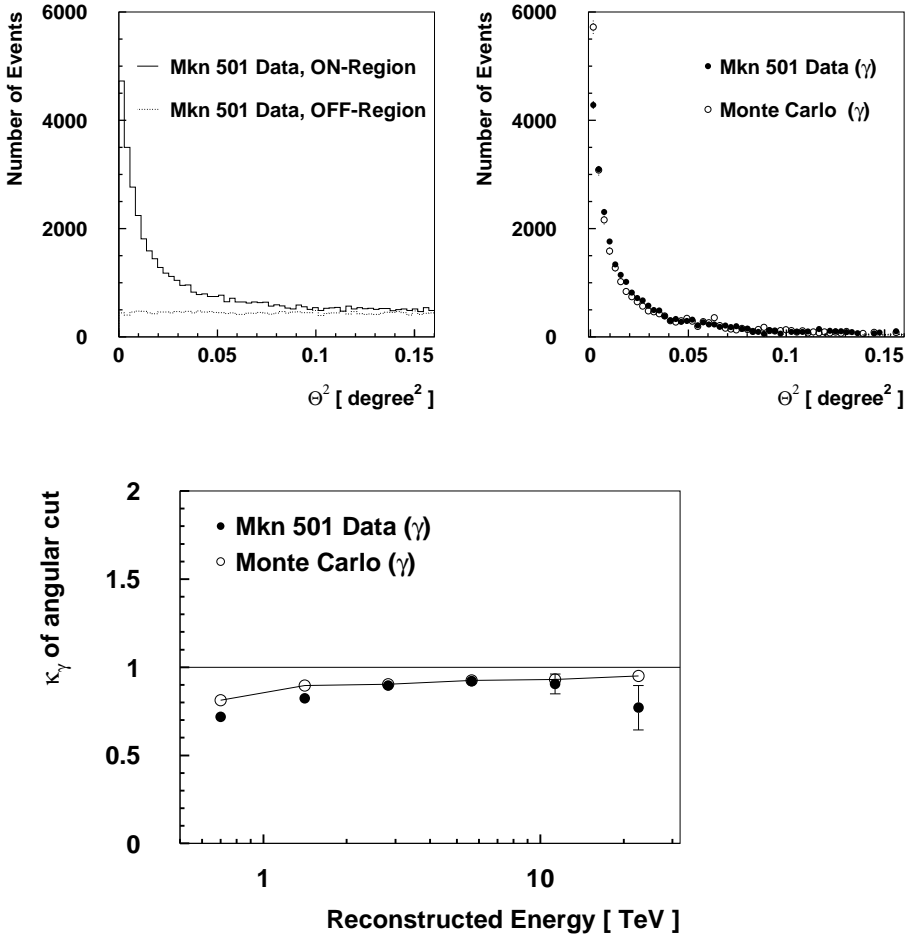}}
\hfill \parbox[b]{55mm}{
\caption{\small \label{angresfig} 
On the left side, the squared angular distances $\Theta^2$
of the reconstructed event directions from the Mkn~501 direction (full line)
and from the OFF source direction (dotted line) are shown
for zenith angles below 30$^\circ$.
On the right side, the measured background 
subtracted $\Theta^2$-distribution (full circles) is compared to the
20$^\circ$ zenith angle Monte Carlo simulations (open circles)
All distributions have been computed using the software threshold of at least 2 IACTs with
$\textsl{size}>40$ and the loose $\gamma$/h-separation cut $\bar{w}_{sc}\,<\,1.3$.}
}
\end{figure*}
\begin{figure*}
\vspace*{0.5cm}
\resizebox{9cm}{!}{\hspace*{1cm}\includegraphics{kappa_theta.epsi}}
\hfill \parbox[b]{55mm}{
\caption{\small 
The $\gamma$-ray acceptance of the cut $\Theta<0.22^\circ$
as a function of the reconstructed primary energy, computed with
the Mkn~501 gamma-rays (full circles) and with the Monte Carlo simulations (open circles).
The computation of the cut acceptances is based on the number of excess events
found in the ON region of angular radius of $0.45^\circ$
(cuts as in Figure \ref{angresfig}). \label{Kappa.t.fig} }
}
\end{figure*}
Based on the stereoscopic images of the shower, 
the shower axis is reconstructed accurately and unambiguously
using a simple geometric method (Aharonian et al.\ \cite{Ahar:97b}). 
The reconstruction permits to 
determine the distances of the telescopes from
the shower axis and consequently, to accurately reconstruct
the shower energy and to efficiently suppress  
the background of cosmic ray-induced showers.

The reconstruction method uses the standard second 
moment parameterization (Hillas \cite{Hill:85}; Fegan \cite{Feg:96}) of the individual images. 
Each image is described by an ellipsoid of inertia computed from the measured
Cherenkov light intensities in the camera.
The intersection of the major axes of {\it two images} superimposed in the ``common
focal plane'', i.e.\  in directional space, 
yields {\it one} estimate of the shower direction.
If more than two telescopes observed a shower,
the arrival directions computed for all pairs of images are combined with a proper
weighting factor to yield the common estimate of the arrival direction.
The weighting factor is chosen proportional to $\sin\,\delta$,
where $\delta$ is the angle between the two major axes.
Taking into account the shower direction, the shower core is
reconstructed using a very similar geometric procedure. 
Note, that this method is based exclusively on the geometry 
of the imaging systems and of the shower axis and does not rely 
on any Monte Carlo predictions.

The angular resolution achieved with this method has been determined
using both Mkn~501 $\gamma$-ray data and the Monte Carlo simulations.
In the case of the Mkn~501 data this is done as follows.
The squares of the angular distances $\Theta$ of
the reconstructed shower directions from the Mkn~501 position are
histogrammed. The subtraction of the corresponding distribution of the fictitious
OFF source yields the background-subtracted distribution 
of the $\gamma$-ray events.
In order to reduce the background-induced fluctuations, the analysis
is performed with the $\gamma$/h-separation cut $\bar{w}_{\textrm{sc}}\,<\,1.3$
(see next subsection for the definition of $\bar{w}_{\textrm{sc}}$). 
In the following, the Monte Carlo photon-induced showers are weighted 
according to a power law spectrum with differential spectral index of $-$2.2.

On the left side of Figure \ref{angresfig} the $\Theta^2$-distributions for the ON and the
OFF regions are shown for the zenith angle interval 0$^\circ$-30$^\circ$.
On the right side of Figure \ref{angresfig} the distribution obtained after
background subtraction is compared to the distribution for the
$20^\circ$ Monte Carlo showers.
There is good agreement between the data and the Monte Carlo. 
The projected angular resolution is 0.1$^\circ$ for showers 
near the zenith and is slightly worse for the 
30$^\circ$- and the 45$^\circ$-showers 
i.e.\  0.11$^\circ$ and 0.12$^\circ$ respectively.
%
%
In the analysis presented in this paper, only a loose cut of 
$\Theta\,<\,0.22^\circ$ is used which accepts, after softwarethreshold, 
~85\% of the photon induced showers and rejects 99\% of the hadron-induced showers. 
By this loose cut the systematic uncertainties caused by the 
energy-dependent $\gamma$-ray acceptance of the
cut are minimized.

In Figure \ref{Kappa.t.fig} the $\gamma$-ray acceptance ($\kappa_{\gamma,\Theta}$) 
of the angular cut is shown as a function of the reconstructed shower energy, 
as determined both from the Mkn~501 and from Monte Carlo data. 
In the case of the Mkn~501 data the same background subtraction technique is used
as described above. 
For the lowest energies ($<$ 1 TeV) the $\gamma$-ray acceptance is slightly lower
than for higher energies, i.e.\  the angular resolution is slightly worse.
This is a consequence of the photon statistics 
per image and the corresponding uncertainty of the images' major
axes.
\begin{figure*}
\resizebox{12cm}{!}{\includegraphics{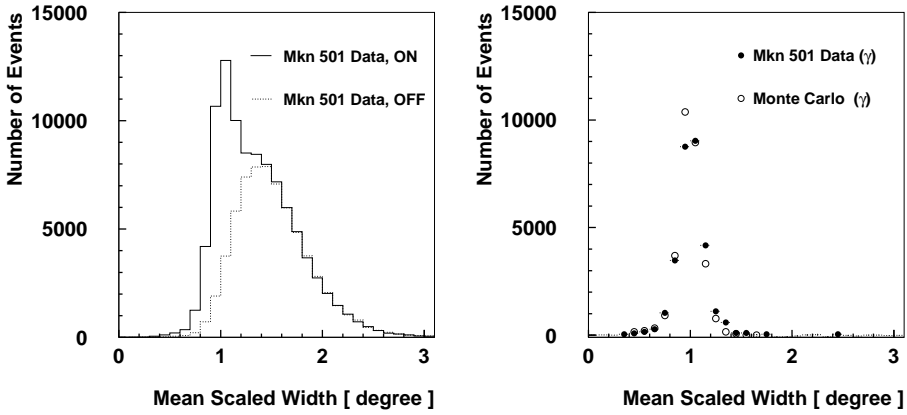}}
\hfill \parbox[b]{55mm}{
\caption{\small \label{imagefig} 
On the left side, the mean scaled width distribution is shown for the ON region (full line)
and for the OFF region (dotted line).
On the right side, the background subtracted distribution (full circles) 
is compared to the Monte Carlo distribution (open circles).  
(all distributions: software threshold: at least 2 IACTs with $\textsl{size}>40$, 
and, cut: $\Theta\,<\,0.3^\circ$, data: all zenith angles, 
Monte Carlo: zenith angle = 20$^\circ$).
}}
\end{figure*}
%
%
\begin{figure*}
\vspace*{0.5cm}
\resizebox{9cm}{!}{\hspace*{1cm}\includegraphics{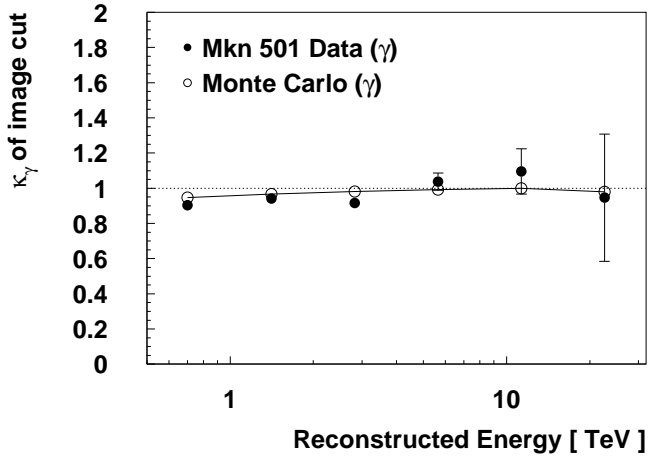}}
\hfill \parbox[b]{55mm}{
\caption{\small 
The $\gamma$-ray acceptance of the cut $\bar{w}_\textrm{sc}\,<\,1.2$
as a function of the reconstructed primary energy, computed with
the Mkn~501 gamma-rays (full circles) and with the Monte Carlo simulations (open circles)
(cuts as in Figure \ref{imagefig}).}
\label{Kappa.w.fig}}
\end{figure*}
At higher energies the angular resolution improves less than expected from the
increase in photon statistics.
This is a consequence of an increasing fraction of showers 
which at higher energies 
are still able to fulfill the trigger criteria, albeit
having impact points far away from the telescope system. 
With increasing distance of the impact point from the telescope 
system, the accuracy of the direction reconstruction
decreases as a consequence of smaller angles between
the image axes of the different telescopes.
\subsection{Image analysis gamma/hadron separation}
The IACT technique permits the suppression of the background of cosmic ray-induced 
showers by the directional information and, additionally, by analysis
of the shapes of the shower images. 
In the case of stereoscopic IACT systems, the gamma/hadron separation
power of the parameter {\small $\textsl{WIDTH}$} 
of the standard second moment analysis, 
which describes the transverse extension of the shower image
in one camera, can be increased substantially, due to two facts:
First, the location and the orientation of the shower axis is 
known from the three-dimensional
event reconstruction and cuts can be optimized accordingly.
Second, a telescope system provides complementary information about the 
transversal extension of the shower obtained from different viewing angles.

The parameter ``mean scaled width'' $\bar{w}_\textrm{sc}$ 
(Konopelko \cite{Kono:95}; Daum et al. \cite{Daum:97}) has been used 
for gamma/hadron-separation. It is defined according to:
{\small
\begin{equation}
\bar{w}_\textrm{sc}\,\,=\,\,\frac{1}{N_\textrm{tel}}\,\,
\sum_{i}\,\,\frac{\textsl{WIDTH}_i}{<\!\textsl{WIDTH}(r_i,size_i,\theta)\!>_
{\gamma}},
\end{equation}}
where the sum runs over all $N_{tel}$ telescopes which triggered.
{\small $\textsl{WIDTH}_i$} is the {\small $\textsl{WIDTH}$}-parameter measured with telescope $i$ and
{\small $<\!\!\textsl{WIDTH}(r_i,size_i,\theta)\!>_{\gamma}$} is the 
{\small $\textsl{WIDTH}$}-value expected for photon-induced showers, 
given the telescope distance $r_i$ from the shower axis,
the total number of photoelectrons, $size_i$, observed in the telescope, 
and the zenith angle $\theta$ of observations.
The {\small $<\!\!\textsl{WIDTH}(r_i,size_i,\theta)\!\!>_{\gamma}$}-values 
are computed from
a Monte Carlo table, using an empirical function for interpolation 
between the simulated zenith angles.
By using a {\it scaled} {\small $\textsl{WIDTH}$}-parameter, it is possible to take into account
that on average the widths of the shower images widen with increasing 
telescope distance from the shower axis and with the total number of 
photoelectrons recorded in a telescope.
By {\it averaging} over the values computed for each telescope, the statistical
accuracy of the parameter determination improves {\it and} 
the information about the shower gained from different viewing angles
is combined.

Fig.\ \ref{imagefig} shows the distribution of the $\bar{w}_\textrm{sc}$ parameter 
for the Mkn~501 $\gamma$-rays and for the Monte Carlo photon data-sample.
The distribution for the Mkn~501 $\gamma$-rays 
has been obtained as follows. 
The $\bar{w}_\textrm{sc}$-values of the events 
satisfying the loose cut on the angular distance $\Theta$ 
from the Mkn~501 location $\Theta<0.3^\circ$ are histogrammed.
The subtraction of the corresponding OFF distribution yields
the background-free distribution of the $\gamma$-ray events. 
As can be seen in Figure \ref{imagefig} 
the experimental distribution and the Monte Carlo distribution
are in excellent agreement.

In the analysis of this paper only a loose cut of $\bar{w}_\textrm{sc}\,<\,1.2$
which accepts, after softwarethreshold, 96\% of the photons and rejects 80\% of the
cosmic ray-induced air showers is used.
With this loose cut the systematic
uncertainties caused by the energy dependent $\gamma$-ray acceptance 
of the cut are minimized. In Figure \ref{Kappa.w.fig}
the $\gamma$-ray acceptances $\kappa_{\gamma,\rm img}$  of the shape cut 
as determined from data and as determined from Monte Carlo 
as a function of the reconstructed energy
are compared to each other. 
The results are in excellent
agreement with each other. 
Due to background fluctuations, 
the determination of the cut acceptance from experimental
data can yield values larger than one.
\subsection{Reconstruction of the primary energy and
determination of differential spectra}
\label{Energy}
The determination of a differential $\gamma$-ray spectrum is performed
in several steps.
In the first step, for all events of the ON and the OFF region the
primary energy $E$ is reconstructed under the assumption that the primary
particles are all photons. 
The reconstruction is based on the fact that,
for a certain type of primary particle 
and a certain zenith angle $\theta$, 
the density of atmospheric Cherenkov light created by the extensive 
air shower at a certain distance from the shower axis
is to good approximation proportional to the energy of
the primary particle.
Presently two algorithms are used:
\begin{enumerate}
\item
The first method is based on the functions \linebreak[4]
{\small $<\!\textsl{size}(r,E,\theta)\!>_\gamma$} and 
{\small $\sigma_{\textsl{size}}(r,E,\theta)^2\,\equiv\,$
$<\!\left(\textsl{size}(r,E,\theta)-\right.$\linebreak[4]
$\left.<\!\textsl{size}(r,E,\theta)\!>_\gamma
\right)^2\!>_\gamma$} which describe the expected sum of 
photoelectrons, $\textsl{size}$, and its variance
as a function of 
the distance $r$ of a telescope from the shower axis,
the primary energy $E$ of the photon, and
the zenith angle $\theta$. 
Both functions are computed from the Monte Carlo event-sample and
are tabulated in $r$-, $E$-, and $\theta$-bins.
Given, for the $i$th telescope, the shower axis distance $r_i$ from
the stereoscopic
event analysis and the sum of recorded photoelectrons $\textsl{size}_i$,
an estimate $E_i$ of the primary energy is made by numerical inversion of 
the function $<\!\textsl{size}(r,E,\theta)\!>_\gamma$.
Subsequently the energy estimates from all triggered
telescopes are combined with a proper weighting factor
proportional to \linebreak[4] $1/(\sigma_{\textsl{size}}(r_i,E_i,\theta))^2$
to yield a common energy value.
\item
In a very similar approach $E$ is estimated from the $\textsl{size}_i$ and the $r_i$
using a Maximum Likelihood Method which takes the full probability density
functions (PDFs) \linebreak[4] $p(\textsl{size};r,E,\theta)$ of the $\textsl{size}$-observable into account.
The PDFs are determined from the Monte Carlo-simula\-tions for
certain bins in $r$, $E$, and $\theta$. The common estimate for the
energy
$E$ maximizes the joint a posteriori probability function $\prod_i
\,p(\textsl{size}_i;r_i,E,\theta)$
that the $\textsl{size}_i$-values have been observed at the zenith angle $\theta$
and at the distances $r_i$.
\end{enumerate}
%
%
\begin{figure}
\resizebox{\hsize}{!}{\includegraphics{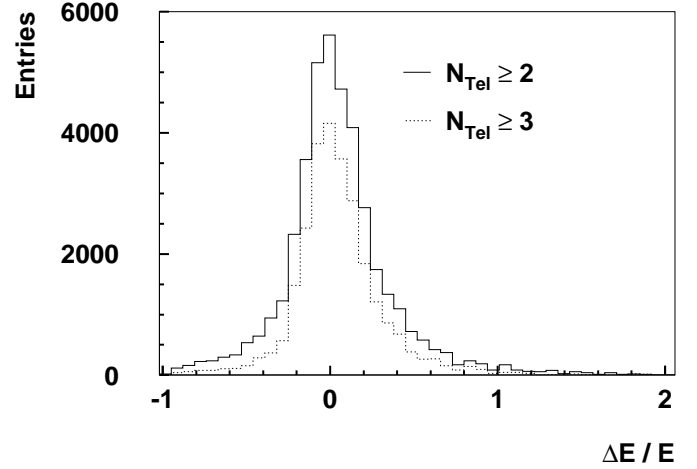}}
\caption{\small The relative error of the energy reconstruction
$\Delta\,E\,/\,E$ (with $\Delta\,E\,>0$ if the reconstructed energy is higher
than the true energy), shown for $\gamma$-ray induced showers 
The full line shows the distribution for all
showers with at least 2 telescopes with $\textsl{size}\,>\,40$ and the dotted line
shows the distribution for all showers with at least 3 telescopes with
$\textsl{size}\,>\,40$ (Monte Carlo, zenith angle 20$^\circ$, 
weighting according to $\mathrm{d}N_\gamma\,/\,\mathrm{d}E\,\propto\,E^{-2.2}$).}
\label{energyfig} 
\end{figure}
Monte Carlo showers have been simulated for the 4 discrete zenith angles
$0^\circ,\,20^\circ,\,30^\circ$, and $45^\circ$.
The energies for arbitrary $\theta$-values between 0$^\circ$ and
45$^\circ$ are determined by interpolation of
the two energy estimates computed with the Monte Carlo tables of the adjacent 
zenith angle values below and above $\theta$.
Hereby an interpolation linear in cos($\theta$)$^{-\zeta}$ is used,
where $\zeta\,=\,2.4$ is derived as described below.
Very small images with $\textsl{size}\,<\,40$ are excluded from the analysis.
Both methods yield the same energy reconstruction accuracy of
$\Delta E/E\, \sim \, 20\%$ for photon-induced showers, 
almost independent of the primary energy.
In the analysis presented below the Maximum Likelihood Method is
used.
%
%
\begin{figure*}
\resizebox{12cm}{!}{
\includegraphics{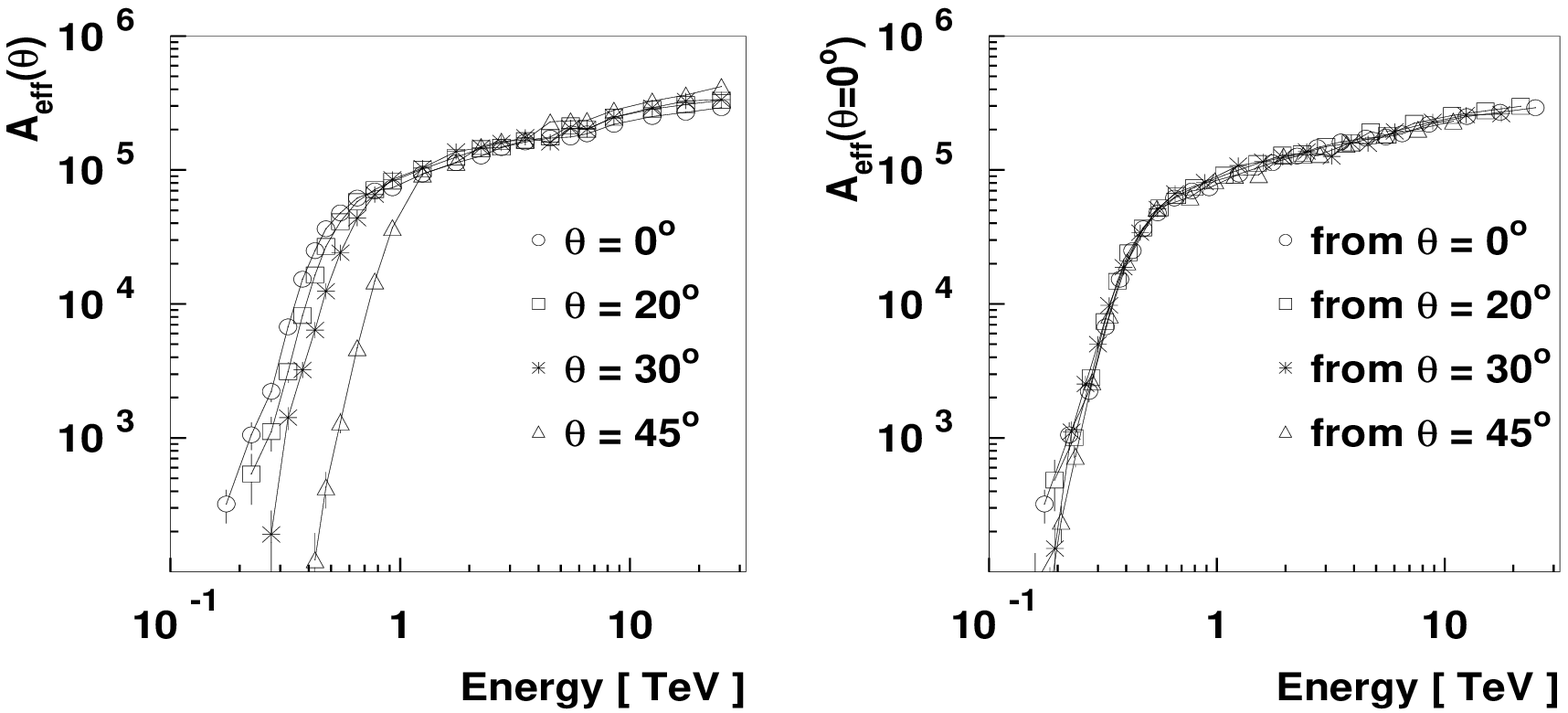}}
\hfill
\parbox[b]{55mm}{
\caption{\small  On the left side the effective areas of the HEGRA system of 4 IACTs
for $\gamma$-ray detection as functions of the primary energy 
are shown for the 4 different zenith angles 
$\theta\,=\,0^\circ,\,20^\circ,\,30^\circ,$\, and $45^\circ$ (Monte Carlo).
The right side shows the effective area for vertically incident
showers $\theta\,=\,0^\circ$ calculated with 
the effective areas at the zenith angles 
$\theta\,=\,20^\circ,\,30^\circ,$\, and $45^\circ$
according to Equation \protect\ref{gl1}
(hardware threshold of at least 2 triggered telescopes, no cuts).}
\label{aeff.fig}}
\end{figure*}
%
%
\begin{figure*}
\resizebox{12cm}{!}{\includegraphics{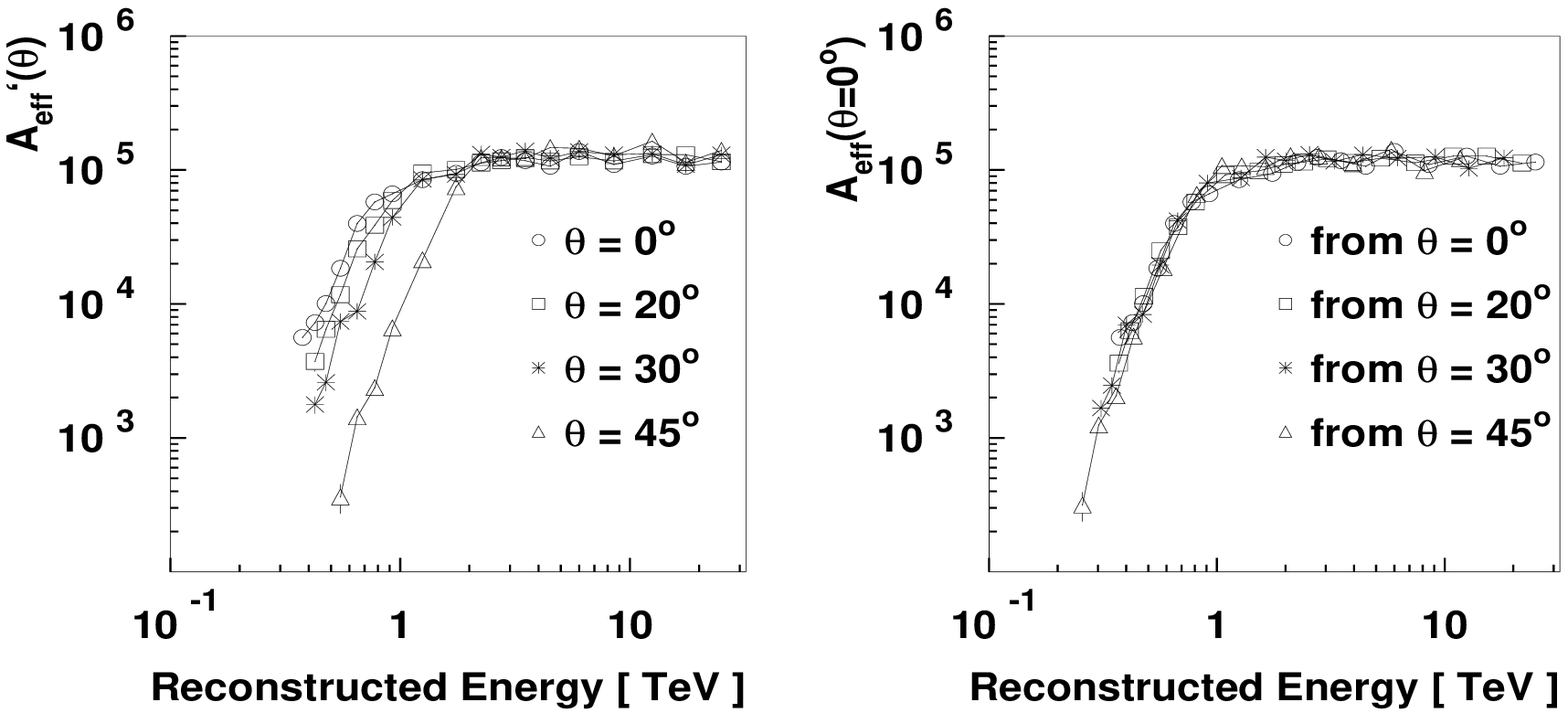}}
\hfill \parbox[b]{55mm}{
\caption{\small 
On the left side the effective areas
as function of the {\it reconstructed} primary energy 
are shown for the 4 different zenith angles (Monte Carlo) .
The same cuts as in the spectral studies have been used, i.e.\ the
software threshold of two telescopes with a $\textsl{size}$-value 
above 40 and a distance of the shower axis from the
center of the telescope system smaller than 195\,m.
On the right side it is shown, how the effective area for vertically incident
showers can be computed from the effective areas 
computed for the other three zenith angles using Equation \protect\ref{gl1}.
}
\label{fit}}
\end{figure*}
In Figure \ref{energyfig} the relative reconstruction error
$\Delta\,E\,/\,E$ is shown for the second method for all triggered
$\gamma$-ray showers with at least two images with $\textsl{size}\,>\,40$, and for all 
$\gamma$-ray showers with at least three images with $\textsl{size}\,>\,40$.
Increasing the requirement on the minimum number of images improves
the energy resolution slightly but reduces the 
$\gamma$-ray statistics.
In the following we are interested in one-day spectra with sparse photon statistics;
consequently we will use the weaker condition of only two telescopes with
$\textsl{size}\,>\,40$.

In the second step, the differential photon flux per energy channel
is computed using the formula
\[
\frac{\mathrm{d}\Phi}{\mathrm{d}E}(E_i)\,=\,
\frac{1}{\Delta t\,\Delta E_i}\,
\]
\[
\mbox{\hspace{.3cm}}
\left\{
\sum_{j=1}^{n_{\textrm{on,{\it i}}}}
\left[\kappa_{\gamma,\Theta}(E_j,\theta_j)\,
\kappa_{\gamma,\rm img}(E_j,\theta_j)\,A_\textrm{eff}'(E_j,\theta_j)\right]^{-1}
\,-\,\right.
\]
\begin{equation}
\label{spectraequ}
\left.\mbox{\hspace{.3cm}}
\sum_{j=1}^{n_{\textrm{off,{\it i}}}}
\left[\kappa_{\gamma,\Theta}(E_j,\theta_j)\,
\kappa_{\gamma,\rm img}(E_j,\theta_j)\,A_\textrm{eff}'(E_j,\theta_j)\right]^{-1}
\right\}
\end{equation}
where $\Delta t$ is the observation time and $\Delta E_i$ is the width of the
$i$th energy bin. The first sum runs over all ON events reconstructed in the $i$th energy
bin. The second sum runs over all OFF events reconstructed in the $i$th energy bin.
$E_j$ is the reconstructed energy of the $j$th event and 
the parameter $\theta_{j}$ is the zenith angle under which the source 
was observed when the $j$th event was recorded.
The second term subtracts on a statistical basis the background
contamination of the ON region.
The effective area $A_\textrm{eff}'$ accounts for the acceptance of
the detector {\it and} its energy resolution. The factors $\kappa_{\gamma,\Theta}$ and
$\kappa_{\gamma,\rm img}$ account for the $\gamma$-ray acceptances 
of the angular cut and the image cut, respectively.
Given the differentail flux, the integral flux can be computed easily by
integrating Equation \ref{spectraequ} over the relevant energy range.

Generally, the effective area is computed from
\begin{equation}
\label{aeffequ}
A_\textrm{eff}(E,\theta)\,=\,
\,\frac{N_{tr}(E,\theta)}{N_{\rm MC}(E,\theta)}
\,A_\textrm{\small MC}(E,\theta)
\end{equation}
where $N_{\rm MC}$ is the number of Monte Carlo $\gamma$-ray-induced showers generated 
for a certain energy and zenith angle bin, $N_{tr}$ is the number
of these showers which trigger the detector and pass the selection cuts, 
and $A_\textrm{\small MC}$ is the area over which the Monte Carlo showers were thrown.
The area $A_\textrm{\small MC}$ is chosen sufficiently large 
(depending on the primary energy and the simulated zenith angle 
between $2\,\times\,10^5$ and $6\,\times\,10^5$\,m$^2$), so that
virtually no exterior events trigger the experiment.

The energy resolution of the detector is taken into account
by using a slightly modified effective area $A_{\textrm{eff},\alpha}'$
which takes, for a given power law spectrum $\Phi(E)\,\propto\,E^\alpha$, 
the response function of the energy reconstruction $p(E,\tilde{E})$ (properly normalized) 
into account:
\begin{equation}
A_{\textrm{eff},\alpha}'(\tilde{E};\theta) \,=\,
\frac{\int\,dE\,\,p(E,\tilde{E};\theta)\,\,A_{\textrm{eff}}(E,\theta)\,\,\Phi(E)}
{\Phi(\tilde{E})}.\end{equation}
In practice, $A_{\textrm{eff},\alpha}'$ is computed with Equation \ref{aeffequ},
weighting the events according to an incident power law spectrum 
with differential spectral index $\alpha$ and
using for $E$ the reconstructed energy and not the true energy. 
Hereby the cut on the distance $r$ of the shower axis from the center 
of the telescope system $r\,<\,$195\,m which is also used in the spectral analysis
is taken into account.

Equ.\ \ref{spectraequ} permits, by definition, to 
reconstruct accurately a differential power law spectrum with index $\alpha$.
Due to the good energy resolution of 20$\%$ of the IACT system,
$A_{\textrm{eff},\alpha}'$ depends only slightly on $\alpha$.
Monte Carlo studies prove that power law spectra with spectral 
indices between $-$1.5 and $-$3 are reconstructed with a systematic 
error smaller than 0.1 using $A_{\textrm{eff},\,\alpha}'$ with 
a fixed value of $\alpha\,=-2.2$.
Furthermore we have tested this method with several other types of
primary spectra, i.e.\  with broken power law spectra 
and with power law spectra with exponential cut-offs.
The method, used with $A_{\textrm{eff},\,-2.2}'$, reproduces the 
input-spectra with good accuracy.

Differential Mkn~501 spectra obtained with this method, as well as
the method for fitting model spectra to the data, will be discussed below.
Alternatively we have tested the standard forward folding technique and
more sophisticated deconvolution methods.
The deconvolution methods yield a slightly improved effective energy resolution at the
expense of a heavier use of detailed Monte Carlo predictions and/or a larger
statistical error of the individual differential flux estimates.

On the left side of Figure \ref{aeff.fig}, $A_{\rm eff}(E,\theta)$ and in 
on the left side of Figure \ref{fit} , $A_{\rm eff}'(E,\theta)$ are shown, computed 
for $\theta_{\rm MC}\,=\,0^\circ,\,20^\circ,\,30^\circ$, and $45^\circ$. 
The zenith angle dependence of the $A_{\rm eff}$-curves
can be described with the following empirical formula:
\begin{equation}
\label{gl1}
A_{\rm eff}(E,\theta)\,=\,
\frac{1}{\cos^\xi(\theta)}
A_{\rm eff}(\cos^\zeta(\theta)\,\cdot\,E,0^\circ)
\end{equation}
with $\xi\,=\,1.7$ and $\zeta\,=\,2.4$ (see Figure \ref{aeff.fig}, right side).
The same formula with the exponents $\xi\,=\,$0.4 and $\zeta\,=\,$2.2
describes the zenith angle dependence of the $A'_{\rm eff}$-curves
(see Figure \ref{fit}, right side).
This formula is used in the data analysis to interpolate $A_{\rm eff}$ 
between the simulated zenith angles.

The Monte Carlo reproduces nicely the following properties
of $\gamma$-ray-induced showers:
\begin{itemize}
\item the shape of the lateral Cherenkov
light distribution as a function of the primary energy
(Aharonian et al.\ \cite{Ahar:98a}),
\item the single telescope trigger probability as function 
of shower axis distance and primary energy, and
\item the distribution of the shower cores,
\end{itemize}
all determined with the Mkn~501 $\gamma$-ray data-sample.
Hence, we are confident that the Monte Carlo correctly predicts the
$\gamma$-ray effective areas, except for a possible scaling factor
$a\,=1.00\pm0.15$ in the energy which derives from the accuracies with 
which the atmospheric absorption and the Cherenkov photon to ADC counts
conversion factor are known. 
Note, that the possible scaling factor $a$ introduces an uncertainty in
the absolute flux estimates, but not in the measured differential spectral indices.

The Crab Nebula is known to be a TeV source with an approximately constant
TeV emission (Buckley et al.\ \cite{Buck:96}). 
We have tested the full analysis chain described above 
and the stability of the IACT system directly with $\gamma$-rays from this source.
Within statistical errors the Crab observations
prove that the IACT system runs stably and 
that the analysis based on the Monte Carlo simulations
accounts correctly for the hardware changes performed 
during 1997 as well as for the 
IACT system's zenith angle dependence of the $\gamma$-ray acceptance
(Aharonian et al.\ \cite{Ahar:98b}).
\section{The 1997 Mkn~501 light curve}
\label{lightcurve}
\begin{figure}
\resizebox{\hsize}{!}{\includegraphics{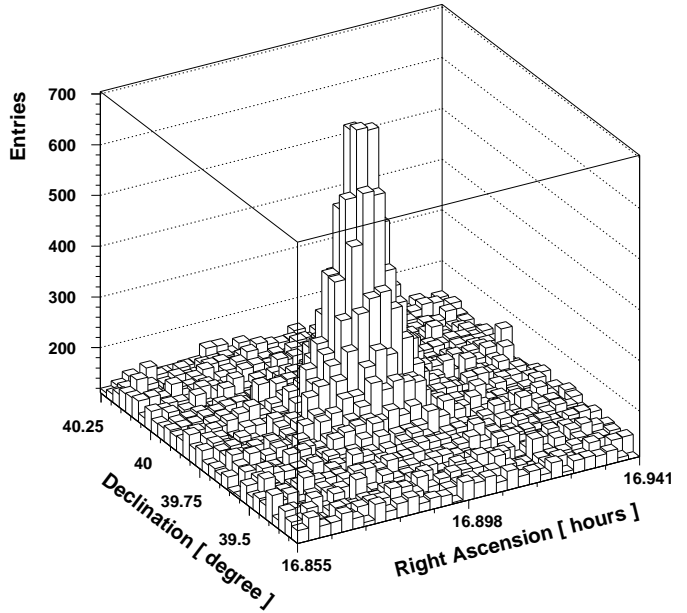}}
\caption{\small 
All events reconstructed in the 1$^\circ$$\,\times\,$1$^\circ$ solid-angle region 
centered on the Mkn~501 direction.
The prominent $\gamma$-ray excess can be seen above the approximately flat
background (hardware threshold, no cuts).
\label{radecfig}
}
\end{figure}

Between March 16th and October 1st, 1997, 110\,h of \linebreak[4]
Mkn~501 data,
satisfying the conditions described above, were acquired.
The excess of about $38,000$ photons is shown in Figure \ref{radecfig} in
equatorial coordinates (hardware threshold, no cuts). 
Note, that the figure shows the number of excess events
as function of declination and right ascension and not a likelihood contour.
The mean location of the excess photons coincides with the Mkn~501 location
with an accuracy of 0.01$^\circ$ (P\"uhlhofer et al.\ \cite{Pueh:97}), corresponding to an
angular
resolution of the IACT system of better than 40 arc sec when limited
by systematic uncertainties and not by statistics.
%
%

%
%
%
\begin{figure*}
\resizebox{\hsize}{!}{\includegraphics{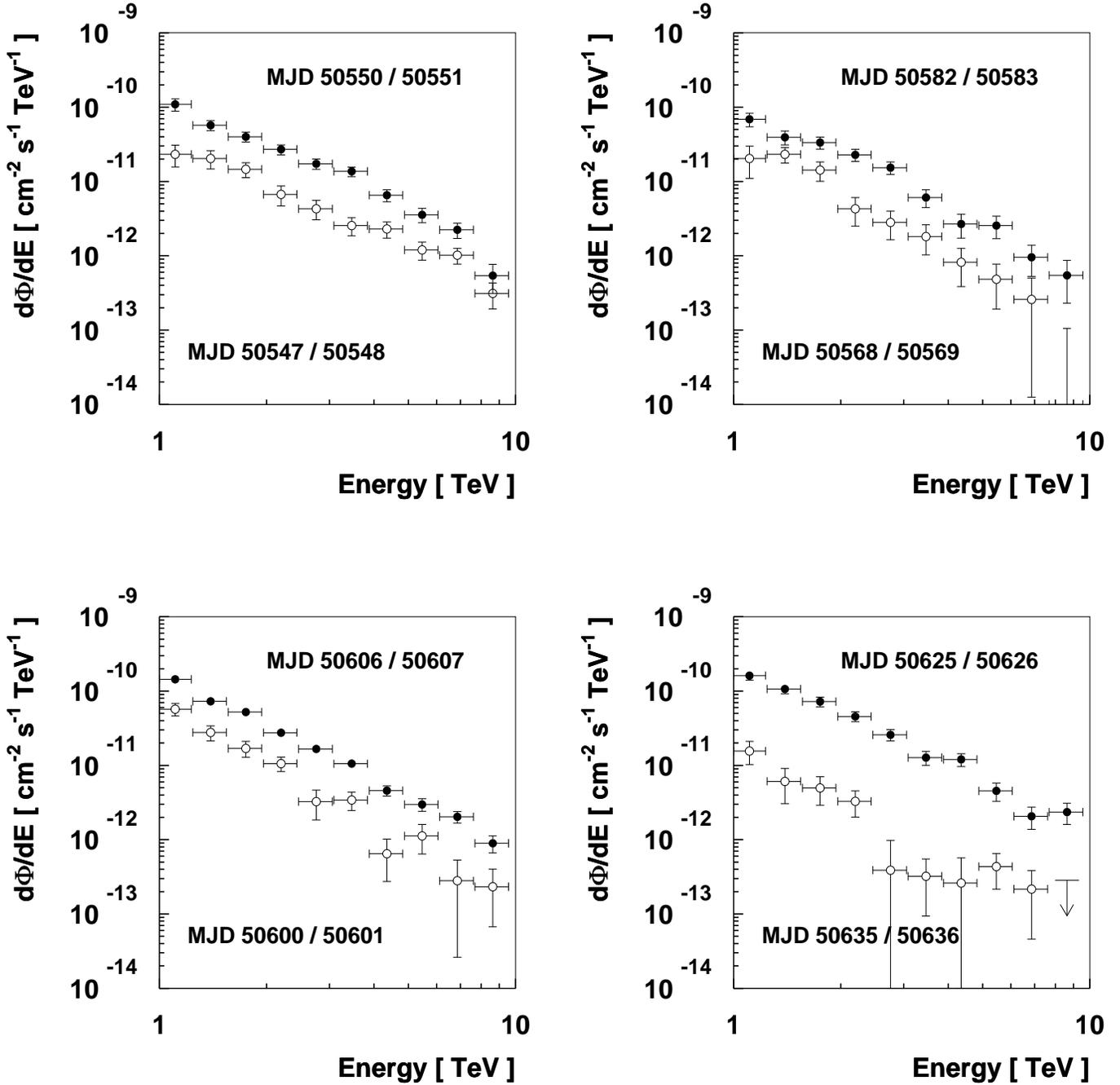}}
\caption{\small \label{spec.02} The differential $\gamma$-ray spectra of eight individual nights.
For each of the four data periods March/April, April/May, May/June, and June/July, 1997
a night of weak emission and a night of strong emission has
been chosen. 
Statistical errors only; see text for systematic errors; the upper limit has 2$\sigma$
confidence level (MJD 50550 corresponds to April 12th, 1997).}
\end{figure*}
%
%
\begin{figure*}
\resizebox{\hsize}{!}{\includegraphics{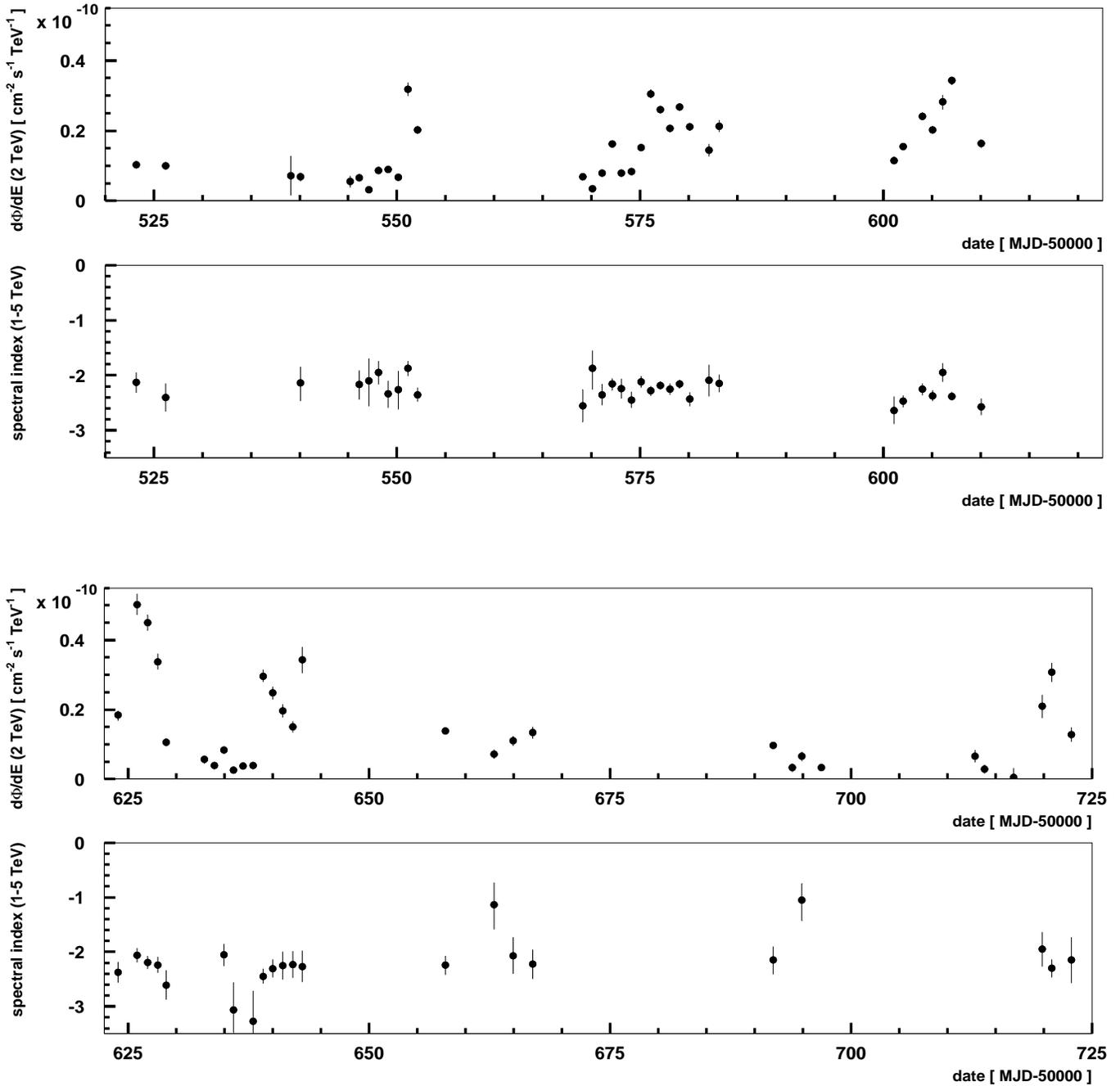}}
\caption{\small \label{fluxfig}
Daily flux amplitudes $\Phi_{2\,\rm TeV}$ and spectral indices (1-5\,TeV) for the
1997 Mkn~501 data. For clearness, spectral indices are shown only for the days with 
sufficient statistics, i.e.\ with errors on the spectral index smaller than 0.5.
Statistical errors only; see text for systematic errors
(MJD 50550 corresponds to April 12th, 1997).}
%
\end{figure*}
%
%
%
\renewcommand{\arraystretch}{1.22}
\begin{table*}
\caption{\small Results on diurnal basis\label{tbl.daily}.
Statistical Errors only; see text for systematic errors.}
\begin{tabular}{llrrr}
\hline
&&&&\\
Start&Duration&$\mathrm{d}\Phi/\mathrm{d}E\,(2\,\rm TeV)$&Spectral index&$\Phi(>1\,TeV)$\\
\footnotesize $\left[ \mbox{ MJD } \right]$
&\footnotesize $\left[ h \right]$
&\footnotesize $\left[\,10^{-12} \mbox{cm$^{-1}$ s$^{-1}$ TeV$^{-1}$}\,\right]$&
(1-5\,TeV)&\footnotesize $\left[\,10^{-12} \mbox{cm$^{-1}$ s$^{-1}$}\,\right]$\\
&&&&\\
\hline
&&&&\\
  50523.1266 &  2.42 &10.22 +0.96{ $-$0.88} & $-$2.13 +0.18{ $-$0.19} & 35.55 $\pm$  3.74\\
  50526.1791 &  1.35 &10.01 +1.05{ $-$1.13} & $-$2.40 +0.25{ $-$0.26} & 36.73 $\pm$  4.14\\
  50527.2305 &  0.32 & $-$ & $-$ &  5.96 $\pm$  8.09 \\
  50539.0783 &  0.32 & 7.21 +5.63{ $-$5.65} & $-$3.48 +3.13{ $-$1.52} & 13.41 $\pm$ 10.02\\
  50540.0758 &  1.21 & 6.79 +1.33{ $-$1.28} & $-$2.14 +0.30{ $-$0.33} & 24.39 $\pm$  5.19\\
  50545.2038 &  0.56 & 5.51 +1.56{ $-$1.58} & $-$2.52 +0.98{ $-$0.68} & 19.21 $\pm$  5.40\\
  50546.0839 &  2.16 & 6.53 +0.83{ $-$0.85} & $-$2.17 +0.26{ $-$0.27} & 22.21 $\pm$  3.16\\
  50547.0566 &  2.68 & 3.16 +0.73{ $-$0.70} & $-$2.10 +0.41{ $-$0.46} & 11.06 $\pm$  2.63\\
  50548.0537 &  2.63 & 8.71 +0.91{ $-$0.83} & $-$1.95 +0.21{ $-$0.22} & 32.79 $\pm$  3.31\\
  50549.0512 &  2.88 & 8.89 +0.93{ $-$0.84} & $-$2.34 +0.24{ $-$0.25} & 34.55 $\pm$  3.89\\
  50550.0561 &  1.75 & 6.66 +1.00{ $-$1.04} & $-$2.26 +0.34{ $-$0.36} & 24.07 $\pm$  3.80\\
  50551.0671 &  1.30 &31.76 +1.95{ $-$1.84} & $-$1.87 +0.13{ $-$0.14} & 111.99 $\pm$  7.22\\
  50552.0751 &  2.27 &20.30 +1.04{ $-$1.18} & $-$2.35 +0.13{ $-$0.13} & 73.99 $\pm$  4.16\\
  50566.0047 &  0.16 & $-$ & $-$ &  6.47 $\pm$ 15.75 \\
  50567.0018 &  1.05 & $-$ & $-$ & 16.14 $\pm$  5.07 \\
  50568.0608 &  0.40 & $-$ & $-$ & 15.20 $\pm$  7.26 \\
  50569.0386 &  1.62 & 6.79 +0.94{ $-$0.94} & $-$2.55 +0.29{ $-$0.30} & 25.38 $\pm$  3.65\\
  50569.9936 &  3.17 & 3.49 +0.64{ $-$0.63} & $-$1.87 +0.32{ $-$0.39} & 11.94 $\pm$  2.73\\
  50570.9909 &  4.08 & 7.89 +0.65{ $-$0.60} & $-$2.35 +0.19{ $-$0.19} & 27.88 $\pm$  2.47\\
  50571.9882 &  4.65 &16.15 +0.66{ $-$0.63} & $-$2.16 +0.10{ $-$0.11} & 57.39 $\pm$  2.80\\
  50572.9855 &  3.68 & 7.89 +0.65{ $-$0.68} & $-$2.24 +0.18{ $-$0.18} & 27.42 $\pm$  2.56\\
  50573.9829 &  4.84 & 8.29 +0.51{ $-$0.56} & $-$2.45 +0.15{ $-$0.14} & 31.70 $\pm$  2.20\\
  50575.0000 &  4.53 &15.21 +0.78{ $-$0.59} & $-$2.12 +0.11{ $-$0.10} & 54.08 $\pm$  2.70\\
  50576.0015 &  2.59 &30.52 +1.24{ $-$1.19} & $-$2.28 +0.09{ $-$0.09} & 109.59 $\pm$  4.78\\
  50576.9745 &  4.80 &26.03 +0.79{ $-$0.77} & $-$2.18 +0.07{ $-$0.07} & 92.66 $\pm$  3.34\\
  50577.9720 &  4.00 &20.71 +0.63{ $-$1.00} & $-$2.25 +0.10{ $-$0.10} & 74.21 $\pm$  3.36\\
  50578.9799 &  3.57 &26.82 +0.81{ $-$1.05} & $-$2.16 +0.08{ $-$0.08} & 95.76 $\pm$  3.96\\
  50580.0128 &  2.24 &21.12 +1.08{ $-$1.22} & $-$2.43 +0.12{ $-$0.13} & 78.77 $\pm$  4.45\\
  50582.0693 &  0.65 &14.47 +1.67{ $-$1.88} & $-$2.09 +0.28{ $-$0.29} & 51.45 $\pm$  6.64\\
  50583.0979 &  0.79 &21.33 +1.77{ $-$1.63} & $-$2.15 +0.16{ $-$0.16} & 79.02 $\pm$  6.16\\
  50601.0393 &  1.30 &11.51 +1.08{ $-$1.09} & $-$2.64 +0.26{ $-$0.24} & 42.92 $\pm$  4.17\\
  50601.9731 &  2.96 &15.52 +0.79{ $-$0.75} & $-$2.47 +0.11{ $-$0.11} & 57.10 $\pm$  2.91\\
  50603.9630 &  2.01 &24.04 +1.23{ $-$0.94} & $-$2.25 +0.10{ $-$0.11} & 87.11 $\pm$  4.16\\
  50604.9625 &  2.60 &20.30 +0.82{ $-$0.99} & $-$2.37 +0.10{ $-$0.10} & 74.36 $\pm$  3.61\\
  50606.0659 &  0.67 &28.19 +2.03{ $-$2.16} & $-$1.95 +0.17{ $-$0.17} & 96.24 $\pm$  7.76\\
\hline
(table continues)
\end{tabular}
\end{table*}
\begin{table*}
\begin{tabular}{llrrr}
\multicolumn{5}{l}{{Table 3---continued.}}\\
&&&&\\ 
\hline
&&&&\\
Start&Duration&$\mathrm{d}\Phi/\mathrm{d}E\,(2\,\rm TeV)$&Spectral index&$\Phi(>1\,TeV)$\\
\footnotesize $\left[ \mbox{ MJD } \right]$
&\footnotesize $\left[ h \right]$
&\footnotesize $\left[\,10^{-12} \mbox{cm$^{-1}$ s$^{-1}$ TeV$^{-1}$}\,\right]$&
(1-5\,TeV)&\footnotesize $\left[\,10^{-12} \mbox{cm$^{-1}$ s$^{-1}$}\,\right]$\\
&&&&\\
\hline
&&&&\\
  50606.9592 &  2.30 &34.39 +1.04{ $-$1.34} & $-$2.38 +0.08{ $-$0.08} & 126.10 $\pm$  4.59\\
  50610.0151 &  1.34 &16.31 +1.18{ $-$1.25} & $-$2.57 +0.15{ $-$0.15} & 60.63 $\pm$  4.52\\
  50623.9119 &  0.98 &18.38 +1.33{ $-$1.57} & $-$2.37 +0.19{ $-$0.19} & 63.91 $\pm$  5.40\\
  50625.9323 &  0.62 &50.20 +3.09{ $-$2.91} & $-$2.06 +0.13{ $-$0.13} & 178.07 $\pm$ 10.12\\
  50626.9801 &  0.87 &44.99 +2.30{ $-$2.18} & $-$2.19 +0.11{ $-$0.12} & 159.90 $\pm$  8.54\\
  50628.0342 &  0.65 &33.72 +2.43{ $-$2.27} & $-$2.24 +0.15{ $-$0.14} & 123.09 $\pm$  8.51\\
  50628.9089 &  0.95 &10.53 +1.22{ $-$1.18} & $-$2.61 +0.27{ $-$0.26} & 38.95 $\pm$  4.43\\
  50631.9110 &  0.94 & $-$ & $-$ & 11.83 $\pm$  3.16 \\
  50632.9106 &  0.98 & 5.68 +1.05{ $-$1.12} & $-$2.24 +0.56{ $-$0.45} & 15.15 $\pm$  3.10\\
  50633.9104 &  0.95 & 3.93 +0.91{ $-$0.93} & $-$2.94 +0.69{ $-$0.52} & 17.37 $\pm$  3.54\\
  50634.9107 &  2.03 & 8.37 +0.78{ $-$0.72} & $-$2.05 +0.20{ $-$0.21} & 28.83 $\pm$  2.79\\
  50635.9088 &  2.09 & 2.54 +0.50{ $-$0.52} & $-$3.06 +0.50{ $-$0.42} & 11.19 $\pm$  2.06\\
  50636.9242 &  0.71 & 3.74 +0.96{ $-$0.99} & $-$3.28 +0.75{ $-$0.54} & 16.29 $\pm$  4.08\\
  50637.9472 &  0.65 & 3.89 +0.95{ $-$0.95} & $-$3.27 +0.56{ $-$0.43} & 19.87 $\pm$  4.10\\
  50638.9759 &  0.89 &29.62 +1.82{ $-$1.72} & $-$2.45 +0.14{ $-$0.13} & 106.12 $\pm$  6.68\\
  50639.9962 &  0.80 &24.77 +1.79{ $-$1.90} & $-$2.31 +0.17{ $-$0.16} & 89.15 $\pm$  6.66\\
  50641.0329 &  0.64 &19.70 +1.85{ $-$2.04} & $-$2.25 +0.25{ $-$0.26} & 72.00 $\pm$  7.52\\
  50642.0453 &  1.06 &15.06 +1.58{ $-$1.69} & $-$2.23 +0.24{ $-$0.25} & 52.92 $\pm$  6.75\\
  50643.0700 &  0.48 &34.39 +3.60{ $-$3.87} & $-$2.27 +0.29{ $-$0.28} & 117.60 $\pm$ 14.86\\
  50657.9012 &  1.55 &13.91 +1.00{ $-$1.06} & $-$2.24 +0.17{ $-$0.18} & 50.27 $\pm$  3.87\\
  50661.0177 &  0.55 & $-$ & $-$ & 25.66 $\pm$  6.73 \\
  50662.9027 &  0.83 & 7.21 +1.33{ $-$1.42} & $-$1.14 +0.41{ $-$0.45} & 23.05 $\pm$  4.42\\
  50664.8955 &  0.82 &10.95 +1.39{ $-$1.42} & $-$2.07 +0.34{ $-$0.33} & 38.63 $\pm$  4.85\\
  50666.9129 &  0.85 &13.36 +1.69{ $-$1.74} & $-$2.22 +0.26{ $-$0.28} & 45.21 $\pm$  6.35\\
  50691.8717 &  1.89 & 9.62 +1.01{ $-$1.00} & $-$2.15 +0.25{ $-$0.26} & 34.64 $\pm$  3.76\\
  50693.8690 &  0.95 & 3.25 +1.26{ $-$1.24} & $-$2.73 +0.93{ $-$0.76} & 13.01 $\pm$  4.36\\
  50694.9192 &  0.67 & 6.53 +1.28{ $-$1.28} & $-$1.05 +0.30{ $-$0.38} & 22.03 $\pm$  6.47\\
  50696.8962 &  1.05 & 3.29 +1.01{ $-$1.03} & $-$1.70 +0.51{ $-$0.62} & 12.50 $\pm$  4.35\\
  50712.8598 &  0.65 & 6.53 +1.76{ $-$1.78} & $-$2.37 +0.74{ $-$0.73} & 24.38 $\pm$  7.22\\
  50713.8566 &  0.90 & 2.89 +1.37{ $-$1.38} & $-$2.56 +1.11{ $-$0.90} &  9.76 $\pm$  4.87\\
  50716.8668 &  0.50 & 0.51 +2.71{ $-$2.70} & $-$1.78 +2.27{ $-$3.22} &  1.76 $\pm$  6.48\\
  50719.8476 &  0.33 &20.91 +3.37{ $-$3.43} & $-$1.95 +0.31{ $-$0.32} & 64.65 $\pm$ 10.40\\
  50720.8479 &  0.65 &30.83 +2.55{ $-$2.92} & $-$2.30 +0.16{ $-$0.17} & 110.59 $\pm$ 11.05\\
  50722.8444 &  0.64 &12.84 +2.07{ $-$2.11} & $-$2.15 +0.42{ $-$0.42} & 48.89 $\pm$  8.37\\
\hline
\end{tabular}
\end{table*}
Fig.\ \ref{spec.02} shows the differential spectra obtained for 8 exemplary individual nights. 
For each of the four data periods March/April, April/May, May/June, and June/July 1997
a night of weak emission and a night of strong emission has
been chosen.
As can be seen, the stereoscopic IACT system, due to its
high signal to noise ratio and due to an energy resolution of $20\%$,
permits a detailed spectral analysis on a diurnal basis.
The spectra can approximately be described by power law models,
although above 5\,TeV the spectra apparently steepen
(see also Aharonian et al.\ \cite{Ahar:97c}; 
Samuelson et al.\ \cite{Samu:98}).

In a first analysis, we fit the data with power laws
in the energy region from 1 to 5 TeV.
The lower bound of the fit region is determined by 
the requirement of the bound being higher than 
the energy threshold of the IACT system in the zenith 
angle interval from 0$^\circ$ to 45$^\circ$.
The upper bound of 5\,TeV has been chosen to minimize systematic
correlations between the emission intensity and the fitted spectral
index which could be caused by the combined effect of a curvature of the 
Mkn~501 spectrum and the effective maximum fit range.
The latter actually is smaller for nights of low Mkn~501 activity, since
for these nights the higher energy channels above $\sim\,$5\,TeV are
frequently not populated and do not contribute to the fit result.

In order to minimize the correlations between the fitted flux amplitude
and the spectral index due to the fitting procedure, the model
$\mathrm{d}\Phi/\mathrm{d}E\,=\,\Phi_{2 \,\rm TeV}\,\cdot\,(E\,/\,2\,{\rm TeV})\,^{\alpha}$
is used, where $\Phi_{2 \,\rm TeV}$ is the differential flux at 2\,TeV and
$\alpha$ is the spectral index.
The energy 2\,TeV approximately equals the median energy of the
Mkn~501 $\gamma$-rays recorded with the IACT system.

We estimate the systematic error on the flux amplitude to be 35\%.
This error is dominated by the 15\% uncertainty of the energy scale.
The systematic error on the spectral index is currently estimated to be 
0.1 and derives from the Monte Carlo-dependence of the results.
The systematic errors as well as more detailed studies of the 
differential Mkn~501 spectra, especially in the energy range below 1\,TeV and 
above 10\,TeV, will extensively be discussed in a forthcoming paper.
Note, that all errors shown in the following plots are statistical only.

The differential fluxes at 2\,TeV and the spectral indices determined on a
daily basis are shown in Figure \ref{fluxfig}. In Table \ref{tbl.daily} the results are
summarized. The gaps in the light curve are caused by the moon, since the IACT system is
only operated during nights without moon.
The emission amplitude is dramatically variable, the measured daily averages
range from a fraction of the Crab emission to $\sim\,$10\,Crab, 
the average flux being 3~Crab. The largest flare was 
observed in June 1997 and peaked at June 26th/27th.

In contrast, the spectral indices are rather
constant. A constant model of $\alpha\,=\,-2.25$ 
fits all spectral indices with a
chance probability for a larger $\chi^2$ of 11\%.
The largest deviations have been found for the nights MJD 50550/50551 
where the spectral index of $-$1.87 +0.13 $-$0.14 deviates by 2.7~$\sigma$
from the mean value and for MJD 50694/50695 where the spectral index
$-$1.05 +0.30 $-$0.38 deviates by 3.2~$\sigma$ from the mean value.
Henceforth, although the IACT system makes it possible to determine the daily
spectral index $\alpha$ 
with an accuracy of $\Delta \alpha$\,$\le$\,0.1 for 15\% of the nights, 
with an accuracy of $\Delta \alpha$\,$\le$\,0.2 for 45\% of the nights,
and with an accuracy of $\Delta \alpha$\,$\le$\,0.35 for 75\% of the nights
the evidence for a change in the spectrum is marginal.

\begin{figure*}
\resizebox{12cm}{!}{\includegraphics{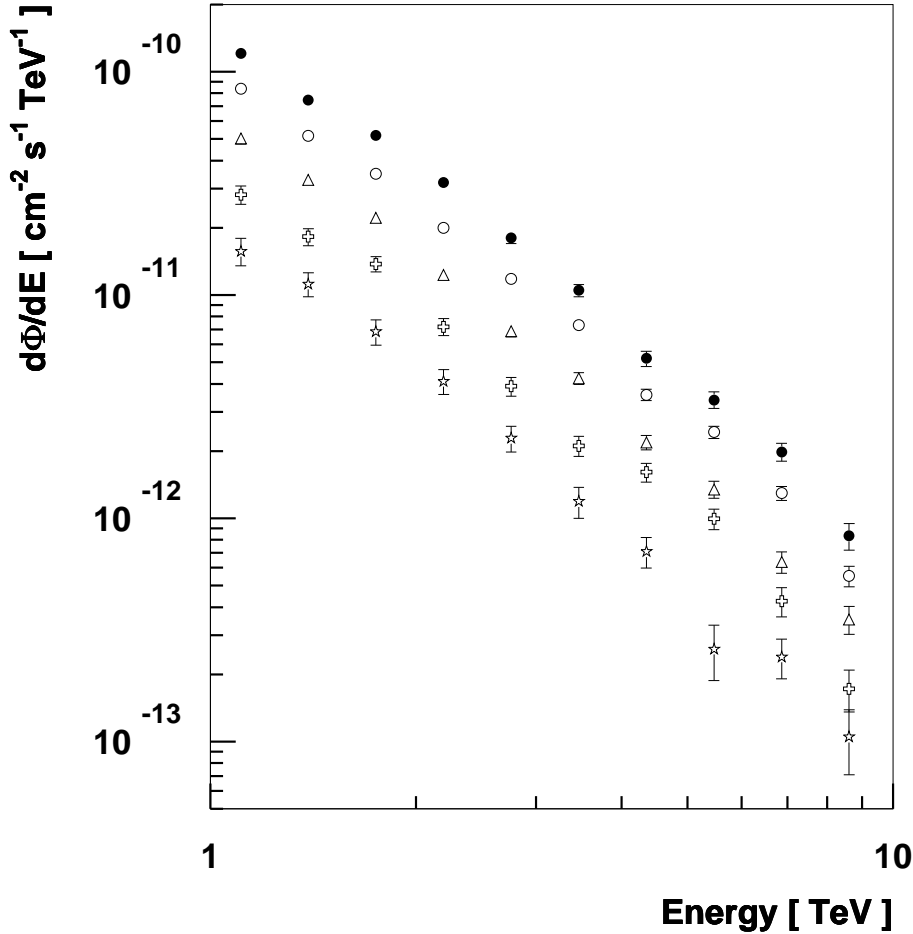}}
\hfill \parbox[b]{55mm}{
\caption{\small \label{highlowfig} 
The Mkn~501 data has been divided into 5 groups according to the activity of the source
as determined on a night to night basis. 
For each of the 5 groups a time averaged spectrum
has been determined. 
The five groups have (from bottom to upwards) $\Phi_{2\,\rm TeV}$ 
$\left[ 10^{-12} \, \rm cm^{-2} \, s^{-1} \,TeV^{-1}\right]$ of below 7,
7--10, 10--20, 20--30, and above 30, respectively.
Statistical errors only; see text for systematic errors.}}
\end{figure*}

\begin{figure*}
\resizebox{\hsize}{!}{\includegraphics{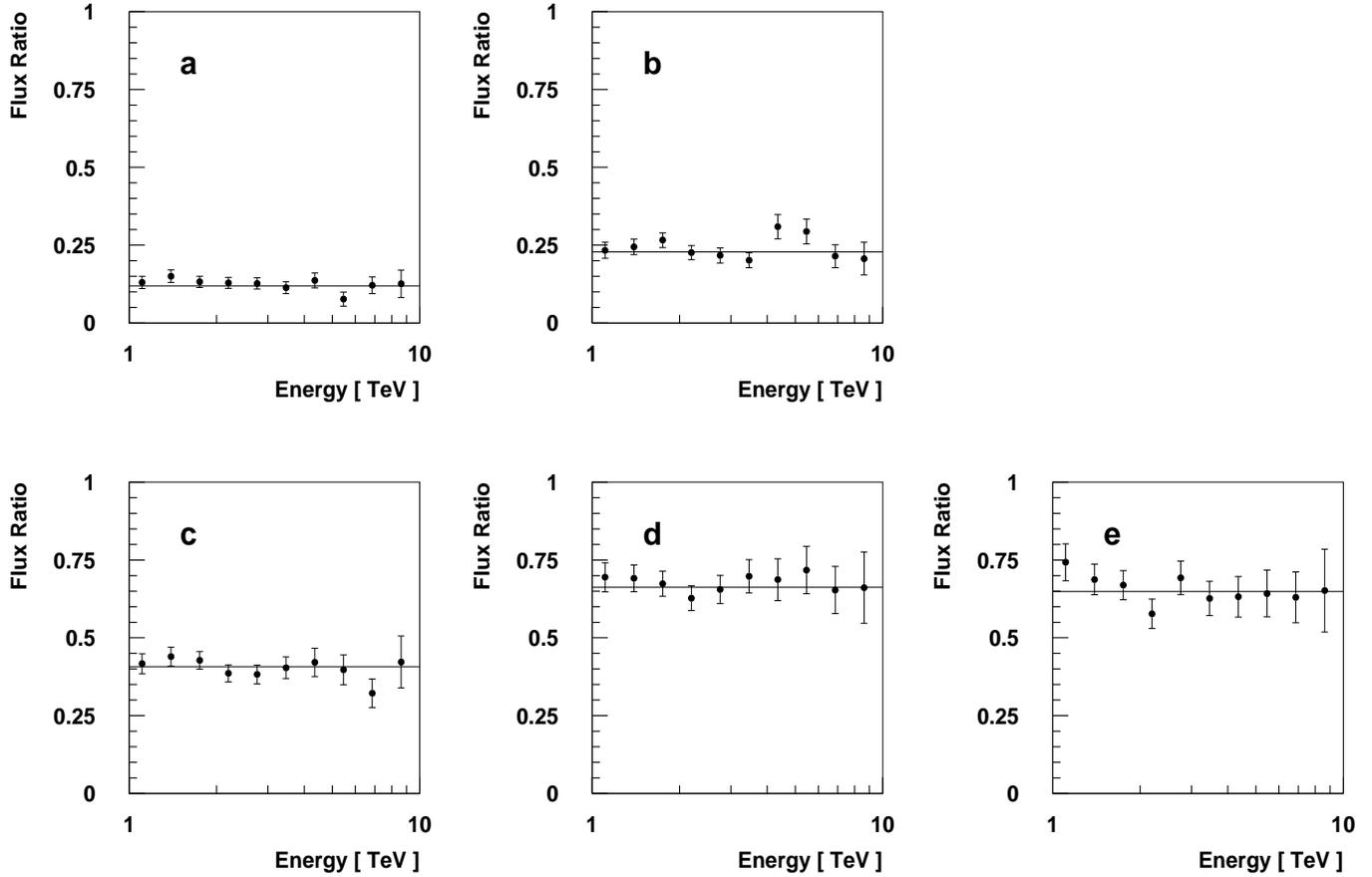}}
\caption{\small \label{div} a-d. The results of dividing the differential spectra
of the four lower flux intensity groups (Figures a-d corresponding to
group i-iv) by the differential spectrum of the 
highest flux intensity group (group v).\protect\linebreak[4] 
{\bf Fig.\ 14.} e. Ratio of the spectrum of data group vii, 
(corresponding to the days following a flare) 
divided by the spectrum of data group vi
(corresponding to the nights at the beginning of a flare 
until a flare reaches its maximum).\protect\linebreak[4] 
The lines show the 
results of constant model fits to the ratio spectra.}
\end{figure*}

In order to study the correlation 
between flux intensity and differential spectrum,
we divided the IACT system data into five groups according to the diurnal 
emission intensity, i.e.\ 
a $\Phi_{\rm 2\,TeV}$-value in units of\linebreak[4]
$\left[ 10^{-12} \,\- \rm cm^{-2} \,s^{-1} \,\- TeV^{-1}\right]$ 
of below $7$ (group i), 7--10 (group ii), 10--20 (group iii), 20--30 (group iv), 
and above 30 (group v), respectively.
In Figure \ref{highlowfig} the differential spectra 
determined with the data of each group are shown.
Within statistical errors the shapes of the five spectra are the same.
This is exemplified in Figure \ref{div}a-d., where the 
four lower flux differential spectra (group i - iv) 
have been divided by the highest flux differential 
spectrum (group v).
For all four cases the ratio of the differential fluxes as a function
of primary energy can be described by a constant model. 
The largest reduced $\chi^2$-values of a constant fit is 1.21
for 9 degrees of freedom, which corresponds to a chance probability 
for larger deviation from the hypothesis of a constant ratio spectrum of 28\%.
In Table \ref{divtab} the power law spectral indices (1 to 5~TeV) 
of the five groups are listed.
Within the statistical errors the first five indices are consistent 
with the mean value of $-$2.25.

%
%
%
%
\begin{table}
\caption{\small \label{divtab} Fit parameters of differential spectra of 7 data groups. Statistical Errors only; see text for systematic errors.}
\begin{tabular}{lllr}
\hline
&&&\\
{Group} 
&{$\mathrm{d}\Phi/\mathrm{d}E\,(2\,\rm TeV)$} 
&{Observation} 
&{Spectral index}\\
&{\scriptsize $\left[ 10^{-12}\rm/cm^{2}\,s\,TeV\right]$} &
time $\left[\rm h \right]$&(1-5\,TeV)
\\
\hline
&&&\\
i   & $<7$ & 26.3 & $-$2.31 +0.10{ $-$0.11} \\
ii  & 7--10  & 23.6 & $-$2.22 +0.06{ $-$0.07} \\
iii & 10--20    & 26.7 & $-$2.27 +0.04{ $-$0.05} \\
iv  & 20--30    & 25.0 & $-$2.24 +0.03{ $-$0.03} \\
v   & $>30$   &  9.5 & $-$2.22 +0.04{ $-$0.04} \\
vi  & ``rising''    & 34.6 & $-$2.19 +0.03{ $-$0.03}\\
vii & ``falling''   & 27.4 & $-$2.29 +0.05{ $-$0.05}\\
\hline
\end{tabular}
\end{table}
We studied whether the TeV spectra during the rising epochs of the light curves
systematically differ from the TeV spectra during the falling epochs of the light
curves.
For this purpose we selected two subsets of the data,
consisting of the accumulated data 
where the flux increased (group vi) or decreased (group vii)
by at least 25\% in comparison to the preceding night.
The ratio of both spectra (group vii divided by group vi) 
as function of energy is shown in Figure \ref{div}e.
A Least Squares fit of a constant model to the ratio 
spectrum gives a reduced $\chi^2$ of 0.72 for 9 degrees of freedom
which corresponds to a chance probability of 69\%.

In Table \ref{divtab} the power law spectral indices (1 to 5~TeV) 
of these two groups are given. There is only a weak indication that group vi
(``rising'' days) with a spectral index of $-$2.19$\pm$0.03
might have a flatter spectrum than group vii (``falling'' days)
with a spectral index of $-$2.29$\pm$0.05.
%
%
%
\section{Study of the shortest time scales of variability}
\label{variability}
The time scales of the Mkn~501 TeV flux variability have been studied
in two ways. First, the time gradients of the daily flux
amplitudes $\Phi_{2 \,\rm TeV}$ have been analyzed.
The analysis uses the exponential increase/decay-constant $\tau$,
as computed for each pair of two daily flux amplitudes
according to:
\begin{equation}
\label{tauequ}
 \tau\,\equiv\,\frac{\Delta\,t}{\Delta\,ln\, \Phi_{2 \,\rm TeV}},
\end{equation}
where $\Delta\,t$ is the time difference between the two measurements and
$\Delta\,ln\,\Phi_{2 \,\rm TeV}$ is the difference between the
logarithms of the differential fluxes at 2\,TeV.
This formula has been derived by assuming a time dependence of
$\Phi_{2 \,\rm TeV}$ according to $\Phi_{2\,\rm TeV}\,\propto\,e^{t/\tau}$.
In the case of small changes in the flux amplitude 
$\Delta\,\Phi_{2\,\rm TeV}\,\ll\,\bar{\Phi}_{2\,\rm TeV}$, where 
$\bar{\Phi}_{2\,\rm TeV}$ is the time averaged flux amplitude,
the ``doubling time'' is commonly used to characterize the variability time scale.
It is defined as the time in which the flux would have increased or decreased by 100\%,
assuming a linear increase or decrease of the flux:
\begin{equation}
\label{doubleequ}
t_{double}\,\equiv
\,\Delta\,t\,\frac{\bar{\Phi}_{2 \,\rm TeV}}
{\Delta\,\Phi_{2\,\rm TeV}}.
\end{equation}
For small changes in the flux amplitude
the increase/decay-constant $\tau$ computed with Equation \ref{tauequ}
equals $t_{double}$ computed with Equation \ref{doubleequ}.

\begin{figure}
\resizebox{\hsize}{!}{\includegraphics{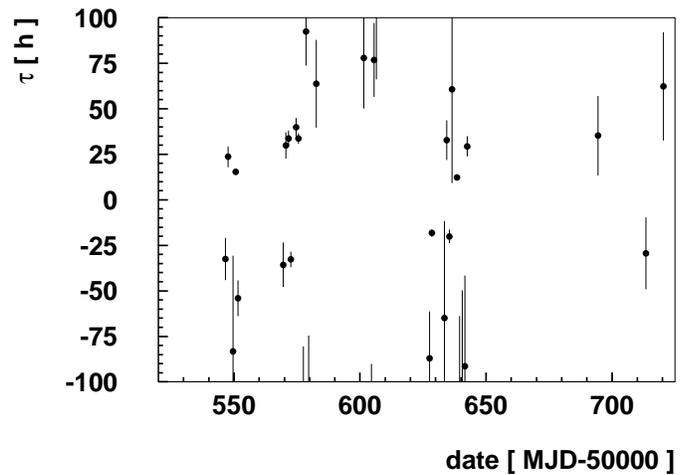}}
\caption{\small \label{varfig} The increase/decay constants $\tau$ computed for the
flux amplitudes $\Phi_{2\,\rm TeV}$ of adjacent nights plotted 
against the mean MJD of the two nights
(MJD 50550 corresponds to April 12th, 1997).
}
\end{figure}

\begin{figure}
\resizebox{\hsize}{!}{\includegraphics{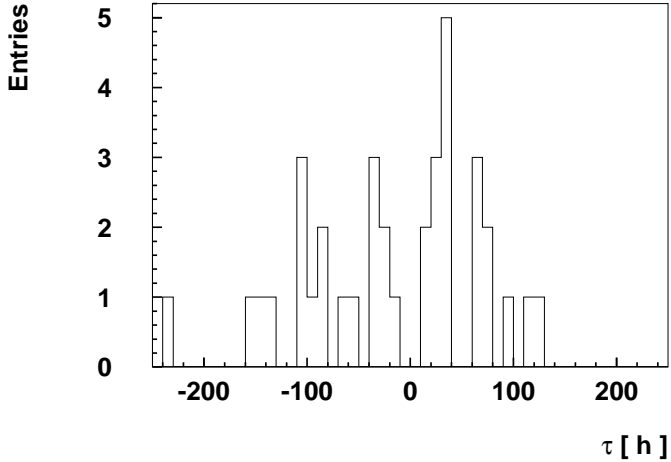}}
\caption{\small \label{var_dis_fig} The distribution of increase/decay constants
$\tau$ computed for the flux amplitudes $\Phi_{2\,\rm TeV}$ of adjacent nights.
Only $\tau$-values between $-$250\,h and +250\,h are shown.
}
\end{figure}

In Figure \ref{varfig} the $\tau$-values computed for adjacent nights are
shown. The most rapid $\tau$-values are in the order of 15\,h. In Table \ref{tbl.tau}
the $\tau$-values more rapid than 24\,h are listed. 
The distribution of the $\tau$-values is to first order approximation symmetric
under time reversal (see Figure \ref{var_dis_fig}), i.e.\  
the daily data does not give obvious evidence for a different rising and
falling behavior.
%
%
%
%

In a second approach a dedicated search for
variability on sub-day time scales has been performed. 
In order to search on time scales well below 
one hour, where the determination of a differential spectrum is plagued
by large statistical uncertainties, the search directly uses  
the $\gamma$-ray excess rates.
Using only data taken under zenith angles below 30$^\circ$ and applying
a cut on the distance $r$ of the shower axis from the center of the telescope array
$r\,<\,200$\,m, the zenith angle dependence of the $\gamma$-ray rate is negligible, 
i.e.\  less than $25\%$ for $\gamma$-ray spectra
with differential spectral indices between $-$2.5 and $-$1.

\begin{figure}
\resizebox{\hsize}{!}{\includegraphics{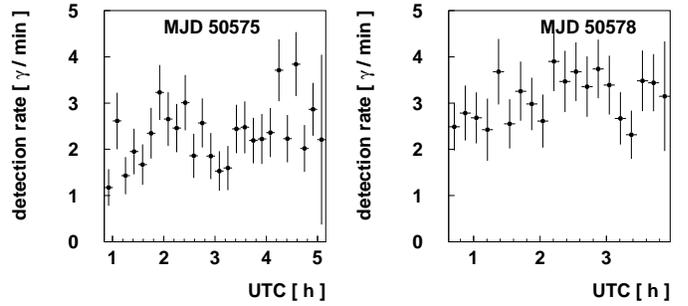}}
\caption{\small \label{coll03} The $\gamma$-ray detection rate 
of two individual nights: 
MJD = 50574/50575 (May 6th/7th), and 50577/50578 (May 9th/10th).
A 10\,minutes binning has been used.}
%
\end{figure}

\begin{table}
\caption{\small The most rapid increase- and decay-times\label{tbl.tau}.}
\begin{tabular}{llrr}
\hline
&&\\
{Night 1 $\left[ \mbox{ MJD } \right]$} 
&{Night 2 $\left[ \mbox{ MJD } \right]$} 
&{$\tau       \,\left[\mbox{ h }\right]$} \\
&&\\
\hline
&&\\
 50547.1      &  50548.1     &  23.6   $\pm$  5.8 \\
 50550.1      &  50551.1     &  15.4   $\pm$  1.6 \\
 50628.0      &  50628.9     &  $-$18.2  $\pm$  2.1 \\
 50635.0      &  50636.0     &  $-$20.1  $\pm$  3.6 \\
 50638.0      &  50639.0     &  12.2   $\pm$  1.5 \\
\hline
\end{tabular}
\end{table}
In Figure \ref{coll03} the excess-rate histories computed with 10\,minute-temporal-bins
are shown for two typical nights, i.e.\  May 6th/7th and May 9th/10th.
Here, as well as in the analysis below, the cuts $\Theta\,<\,0.13^\circ$ and 
$\bar{w}_\textrm{sc}\,<\,1.2$ have been applied.
No strong variability can be seen and a method is needed to decide
on the statistical significance of the observed rate variations.

For the systematic search, a method introduced in (Collura \-and \-Rosner \cite{Coll:87}) has been used.
It is based on the $\chi^2$-fitting
technique and permits a search on multiple time scales. It
provides easy and straightforward
computation of the significances of the detected variabilities.
For each night a variability is searched for, using different time bin sizes, by 
computing the $\chi^2$-value of a constant 
model fit to the data, and performing an averaging procedure
over the relative location of the data points within these binning schemes.
We use time bins with durations $\Delta T$ between
10\,minutes and 2.24 hours, i.e.\  with $\Delta T\,= \,(2^{n/4}\,\times\,10$\,minutes) for $n\,=\,1$ to 15.
The lower limit on the bin duration is given by the requirement of an expected
number of recorded events per bin 
{ \raisebox{-0.5ex} {\mbox{$\stackrel{>}{\scriptstyle \sim}$}}}$\,$5. 
The upper limit is determined by the maximum duration
of the Mkn~501 observations per night which is in the order of 4\,h.
For each night the bin duration which yields the most significant variability
detection is determined and the chance probability $p_{c}$ for a constant 
flux to yield a more significant variability is computed
using an analytic approximation.

%
%
%
%
\begin{figure}
\vspace{1.2cm}
\resizebox{\hsize}{!}{\includegraphics{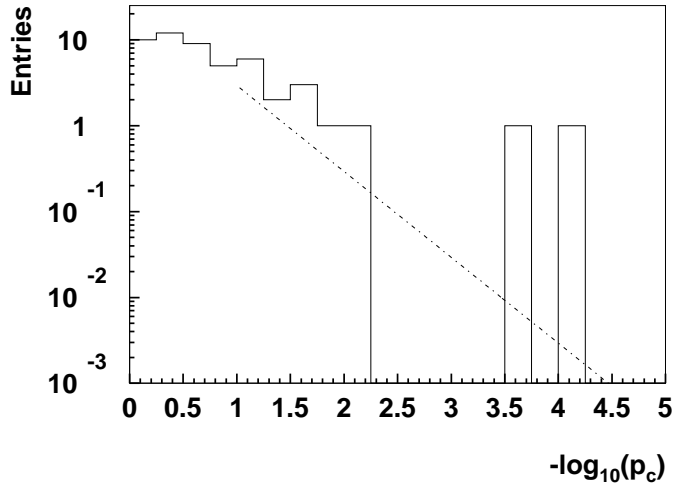}}
\caption{\small \label{coll01} The results of the search for sub-day variability.
For each night, the chance probability $p_c$, computed with an analytic approximation,
with which a constant signal would yield a more significant variability as the observed one
has been computed.
Shown is the distribution of the $-$log$_{10}$($p_c$)-values (histogram) together with 
distribution expected for small chance probabilities 
in the absence of variability (dotted line).
}
\end{figure}
%
%
\begin{figure}
\vspace*{0.5cm}
\resizebox{\hsize}{!}{\includegraphics{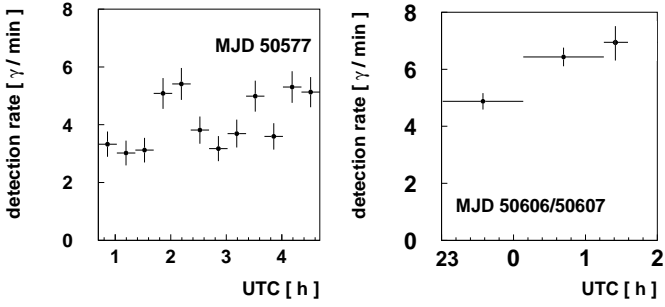}}
\caption{\small \label{coll02} The $\gamma$-ray detection rate 
of the two nights: MJD = 50576/50577 (May 8th/9th) and
50606/50607 (June 7th/8th).
For the first night, a 20\,minutes binning
and for the second night, a 67\,minutes binning has been used.}
%
\end{figure}
The chance probabilities, actually the $-$log$_{10}$($p_c$)-\linebreak[3]
values, for the 51 nights for which sufficient data with zenith angles smaller than 30$^\circ$
is available, are shown in Figure 
\ref{coll01} together with the distribution of large $-$log$_{10}$($p_c$)-values
expected in the absence of flux variability. 
The $-$log$_{10}$($p_c$)-values distribute as expected in the absence of any
variability, except for values near 0 (chance probabilities near 1) 
and except for values larger than 3 
(chance probabilities smaller than 1 per mill).
The deviation at small values derives from the analytical computation
of the chance probabilities, and is thus a pure artifact of the method
(see also Collura \-and \-Rosner \cite{Coll:87}).
The deviation at large values correspond to two nights, 
the night of May 8th/9th (MJD 50576/50577) and the night of June 7th/8th, 1997
(MJD 50606/50607), for which sub-day variability is indicated.

In the case of the first night,
the most significant variability has been found with
20\,minute bins; in the case of the second night with 67\,min
bins. 
The excess rates during the two nights are shown in Figure \ref{coll02}.
In the night from May 8th to May 9th, 1997, the $\gamma-$ray rate oscillated
between 3\,$\gamma$/min and 5.4\,$\gamma$/min 
on a time scale of 1.5 hours.
In the night from June 7th to June 8th, 1997, the $\gamma-$ray rate
continuously increased from 4.9\,$\gamma$/min to 7\,$\gamma$/min within 2.6\,h.
The chance probability for a more significant apparent variability at constant flux
is computed to be $0.8\,\times\,10^{-4}$ for the first night and
$2.0\,\times\,10^{-4}$ for the second night.
Taking into account that 51 nights have been searched for variability, 
the chance probabilities increase to 0.4\% for the first night and to 
1\% for the second night. 
The variability detected in these two nights corresponds to
an increase/decay constant of about 3~h for the first night and
7~h for the second night.

To summarize, the study of the time gradients of the diurnal fluxes yields
smallest increase/decay times of the order of 15\,h. The dedicated search for
flux variability within individual nights on time scales between 10\,minutes and several
hours yielded weak evidence for variability on time scales of around 5 hours.
\section{Correlation of the TeV emission with the X-ray emission}
\label{correlation}
A correlation of X-ray and TeV fluxes would give 
clues regarding the emission mechanism. 
The All Sky Monitor on board the {\it Rossi X-Ray 
Timing Explorer} has been regularly observing bright X-ray sources in the
energy range from 2~-~12\,keV since January 5th, 1996. It observed
mainly X-ray binaries and a list of initially 10, and since May 1997 
a list of 74 bright active galactic nuclei. Each object is monitored roughly 5 times a
day, each time for 90 seconds. The detection threshold per 90 second observation is 30\,mCrab.
The ASM data is publicly available over the Internet.

\begin{figure*}
\resizebox{12cm}{!}{\includegraphics{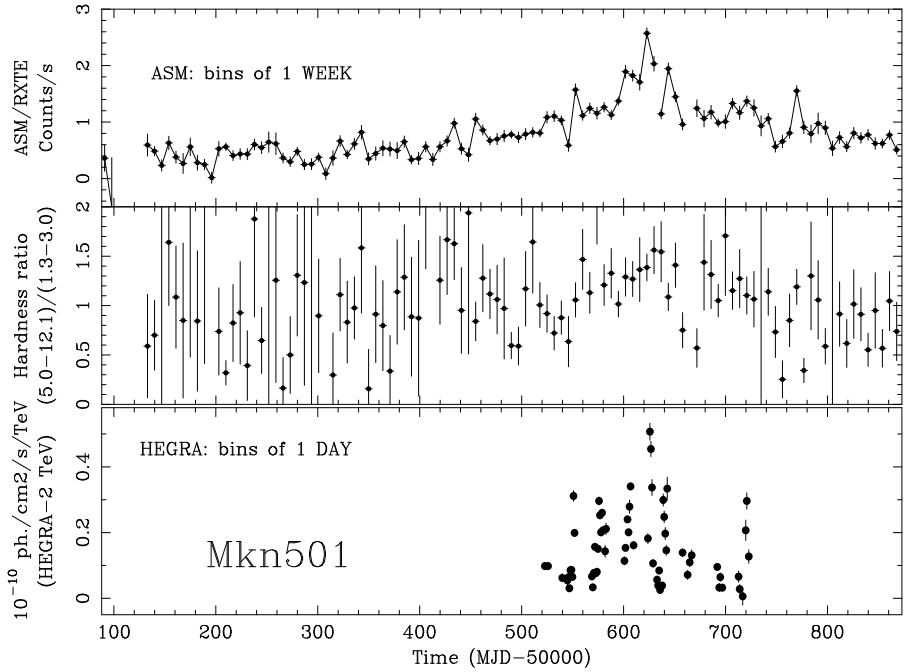}}
\hfill \parbox[b]{55mm}{
\caption{\small \label{asmallfig} The ASM count rates $r_{\small \rm X}$ (2-12\,keV), the ASM hardness ratios 
{\it counts}(5-12.1\,keV)\,/\,{\it counts}(1.3-3.0\,keV), and the 
daily HEGRA differential fluxes at 2\,TeV against time, 
for the time period from January 1997 until February 1998.}}
\end{figure*}

The ASM monitors Mkn~501 since January 5th, 1996. 
In Figure \ref{asmallfig} the time histories of the 
ASM flux and the hardness ratio
{\it counts}(5-12.1\,keV)\,/\,{\it counts}(1.3-3.0\,keV), \linebreak[4] both
computed with bins of 1 week duration, are compared to
the light curve of the HEGRA IACT-system.
We derived the ASM count rates $r_\textrm{\small X}$ from the "definitive" ASM products 
satisfying the requirement of a dwell duration larger than 30\,s 
and a flux fit with a reduced $\chi^2$-value smaller than 1.25.
We excluded days with poor sampling (less than 25\% of the data),
by using only the diurnal rates values which have
an error smaller than 0.375 counts/sec. 
The binned light curves, hardness ratios, and correlation coefficients 
("slow" method with error propagation) have been obtained using the 
``ftools 4.0'' package.

The count rate increases from 0.4/sec in February, 1996, 
slowly to 1\,counts/sec in January, 1997, and then dramatically to
2\,counts/sec in June/July 1997. 
After \linebreak[4] reaching its maximum of 3.1$\pm$0.4 on June 24th, 1997,
the count rates returned to around 1\,counts/sec until April 1998.
During the major flaring phase in 1997, the X-ray spectrum hardens, i.e.\ 
the hardness ratio increases from 0.8 in January 1997
to 1.5 in July 1997 and decreases again to 1 until April 1998.

A correlation between the X-ray activity and the TeV-activity can be recognized
in the sense that the X-ray activity peaked in June/July when the
amplitudes of the TeV flares reached their maximum.

\begin{figure}
\resizebox{\hsize}{!}{\includegraphics{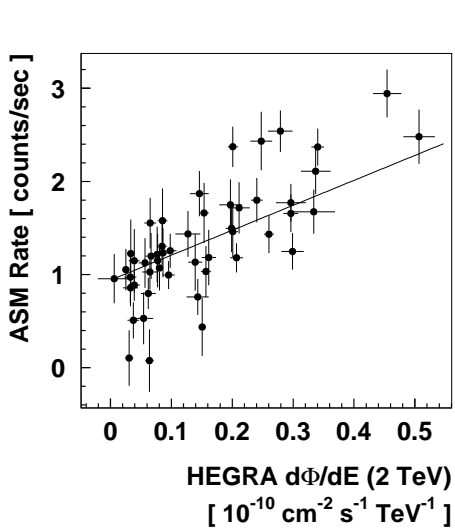}}
\caption{\small \label{2dcorrfig} The correlation of the one-day ASM count rates $r_{\small \rm X}$ (2-12\,keV)
with the daily HEGRA differential fluxes at 2\,TeV. Superimposed is a straight
line fit to the data.}
\end{figure}

In Figure \ref{2dcorrfig} the correlation between the daily differential flux
at 2\,TeV, $\Phi_{2\,\rm TeV}$, and the count rate $r_{\small \rm X}$ is shown. 
Hereby, for each daily $\Phi_{2\,\rm TeV}$-value, 
the ASM rate has been averaged over all 90 second measurements
within the 24\,h time interval centered close to 0:00\,UT.
One sees indications of a correlation between the emission in the
two energy bands. 
A fit to the data gives the correlation:
\[
r_{\small \rm X} \,
\left[ \textrm{counts/sec} \right]
\,=\,
0.94
{\raisebox{-0.5ex}{$\stackrel{+{0.05}}{\scriptstyle-{0.06}}$}}
\,+\,(2.7\,\pm\,0.3)\, 
\]
\begin{equation}
\mbox{\hspace{.3cm}}
\times \, \frac{\Phi_{2\,\rm TeV}}
{10^{-10}\,\mbox{cm$^{-2}$ s$^{-1}$ TeV$^{-1}$}}.\end{equation}
%

%
%
%
%
%
%
\begin{figure}
\resizebox{\hsize}{!}{\includegraphics{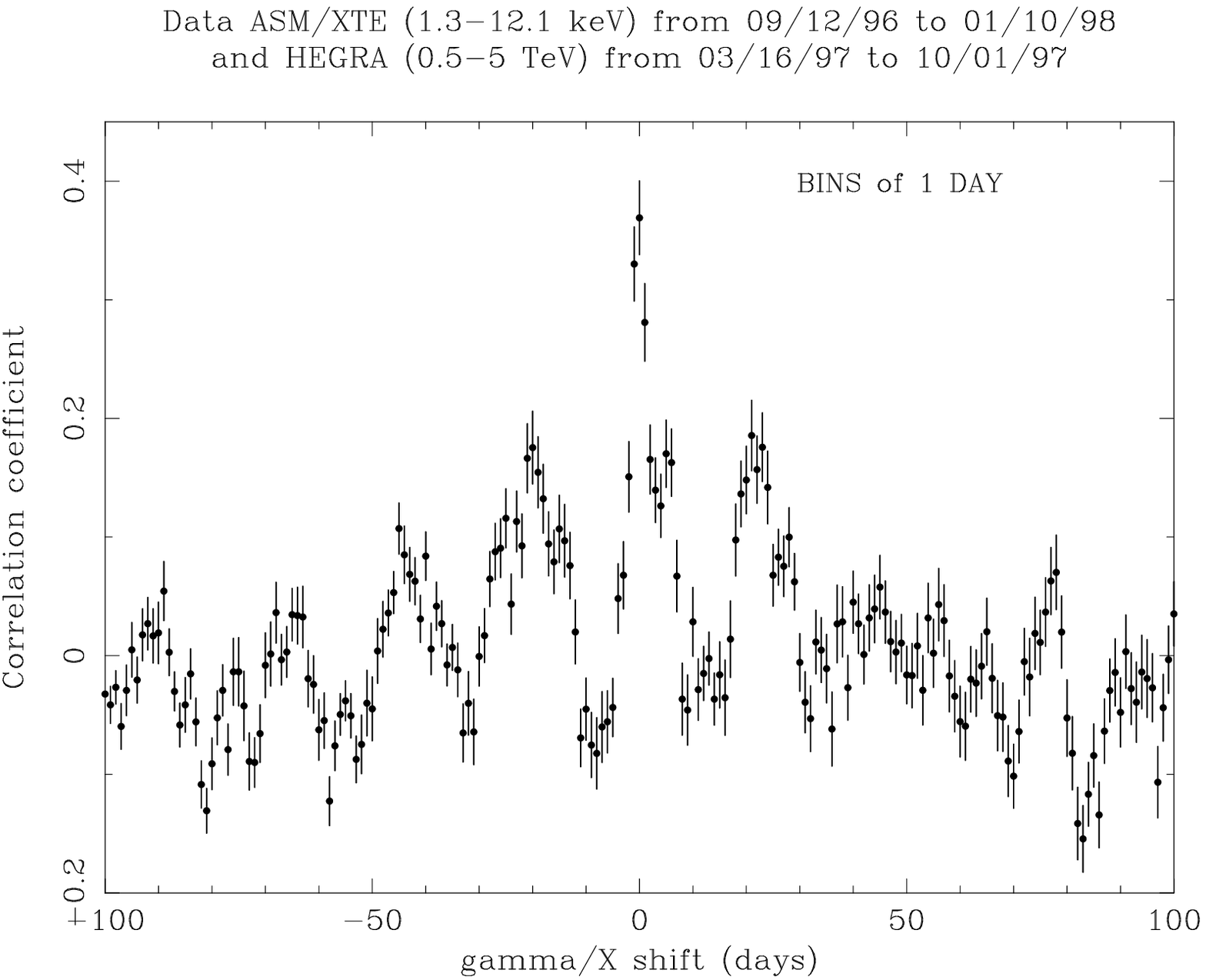}}
\caption{\small \label{corrfig} 
The correlation coefficient of the 
daily HEGRA differential fluxes at 2\,TeV
and the one-day ASM count rates $r_{\small \rm X}$ (2-12\,keV)
as a function of the time shift $\Delta t$ between the 
considered TeV  and X-ray fluxes.
Positive $\Delta t$-values correspond to the TeV variability preceding
the X-ray variability.}
\end{figure}
A possible time shift $\Delta t$ between the TeV- and X-ray variability has
been searched for by computing the discrete correlation coefficient, DCF, 
(Edelson \& Krolik \cite{Edel:88})
\[ \xi\,(\Delta t)\,=\, \]
\begin{equation}
\mbox{\hspace{0.03cm}}
\frac{\sum_i\, (\Phi_{2\,\rm TeV}(t_i)-\overline{\Phi_{2\,\rm TeV}})(r_{\small \rm X}(t_i+\Delta t)-\overline{r_{\small \rm X}})}
{\sqrt{\sum_i\,(\Phi_{2\,\rm TeV}(t_i)-\overline{\Phi_{2\,\rm TeV}})^2\,\,
\sum_i\,\,(r_{\small \rm X}(t_i+\Delta t)-\overline{r_{\small \rm X}})^2}}
\end{equation}
as function of $\Delta t$.
The index $i$ runs over all nights with TeV-measurements, the $\Phi_{2\,\rm TeV}(t_i)$
are the daily TeV flux amplitudes, and the $r_{\small \rm X}(t_i)$ 
are the ASM count rates averaged over 24\,h, centered close to 0:00\,UT.
In Figure \ref{corrfig} the results are presented.
The DCF shows a positive peak reaching from
$\Delta t\,=\,-1$
(TeV variability follows X-ray variability after 1 day) to $\Delta t\,=\,1$ 
(TeV variability precedes X-ray variability by one day).

The data indicates a correlation of the TeV- and X-Ray emission with a time 
lag of one day or less. 
For $\Delta t\,=\,0$, 50 pairs of TeV and X-ray data 
enter the calculation and give a DCF of 0.37$\pm$0.03.
Even completely uncorrelated time series are expected to produce 
non-zero values of the DCF (Edelson \& Krolik \cite{Edel:88}).
The probability distribution of the DCF depends on the 
number of pairs used for its calculation and on the
temporal autocorrelation characteristics of the TeV emission and the 
X-ray emission. 
Assuming 50 statistically independent flux measurements in two energy bands 
which follow Gaussian distributions around their mean values, 
the chance probability for DCF-values exceeding 0.37$\pm$0.03 
is 0.43\%. Reducing the number of statistically independent 
flux pairs from 50 to 15, increases this chance probability to 8\%. 
The true chance probability of the correlation indicated in Figure
\ref{2dcorrfig} and Figure \ref{corrfig} will lie between 
these two extremes. 
Note, that the same correlation is found in the CT1 data and in the CT2 data 
(see Part II). 

We interpret the structure of the DCF which can be recognized in Figure \ref{corrfig}
as to arise from the periodic gaps in the HEGRA observation time, paired with the
spiky structure of the Mkn~501 light curve in both energy bands.
\section{Discussion}
\label{Discussion}
The observations of Mkn~501 during its remarkable state of flaring activity
in 1997 with the HEGRA IACT system allowed us to study in detail
the temporal and spectral characteristics of the source 
with, for gamma-ray astronomy, unprecedented photon statistics and precision. 
More than 38,000 TeV $\gamma$-ray photons were detected 
during March 1997 until October 1997. 
These photons enabled the localization of the $\gamma$-ray 
source with an accuracy of about 40 arc seconds.

The mean flux of $\gamma$-rays averaged over the whole period of activity 
was as high as three times the flux of the  Crab Nebula,
the strongest persistent TeV source in the sky. 
For a source of this strength, even ``loose'' shape cuts 
result in an almost background free detection of $\gamma$-rays:
several 100 $\gamma$-rays against 20 background events caused by
cosmic rays. 
This implies the statistically significant detection of the source 
every few minutes during the whole 6 months of 
observations and makes it possible to
study the flux variability on sub-hour time scales. 
Furthermore, the good precision of reconstruction of the energy of
{\it individual} $\gamma$-rays with $20 \%$ resolution
combined with high $\gamma$-ray statistics makes it possible to  
measure the energy spectra of the radiation and their evolution in time 
on a night-by-night basis.

In this paper we presented the results obtained from IACT system data.
The data of CT1 and CT2 are described in Part II.
The IACT system data has not given any evidence for 
a correlation between the emission
intensity at 2\,TeV and the spectral index, determined between 1 and 5\,TeV. 
The study of the time gradient of the diurnal flux at 2\,TeV yielded shortest
increase/decay times of the order of 15\,h. A dedicated search for short time 
variability within individual nights yielded evidence for a variability with a
corresponding increase/decay time of the order of 5\,h.
The data indicated a weak correlation between the TeV-flux amplitudes and 
the 2 to 12 keV X-ray flux, fovouring a time lag 
between the TeV- and the X-ray variability of one day or less. 
In the following we will briefly discuss these results.

\subsection{Spectral characteristics}
Commonly it is believed that the study of the TeV $\gamma$-ray spectra of 
BL Lac objects at different epochs of their activity provides key 
insights into the nature  of the $\gamma$-ray production processes 
in the relativistic jets. 
Generally, in these highly dynamical objects, when the flux could be changed
by an order of magnitude within 1 day of observations, strong
time-variation of the spectral shape of the radiation is expected as well. 
However, the average spectra of Mkn~501 in the energy range from 1\,TeV to 10\,TeV 
corresponding to largely  different absolute flux levels, 
appear to be very similar as discussed in Section 
\ref{lightcurve} (see Figure \ref{highlowfig}).

In the framework of Inverse Compton models 
this could be interpreted as result of (1) a time-independent
spectrum of accelerated electrons, 
together with (2) a very fast radiative cooling
of the electrons which establishes an equilibrium spectrum of electrons
during the time required for the collection of sufficient photon statistics
for proper spectral measurements (typically a few hours or less if the absolute
$\gamma$-ray flux exceeds the flux level of the Crab). 
At first glance, this contradicts the observed dramatic  
shift of the synchrotron peak in the Mkn~501 spectrum
by 2 orders of magnitude in frequency, discovered with BeppoSAX
during the April 1997 flaring phase (Pian et al.\ \cite{Pian:98}).
Formally speaking, the position of the synchrotron peak $\nu_{\rm s}$ 
is proportional to $B \, \delta_{\rm j} \, E_{\rm max}^2$, 
henceforth its variation could be explained by the variation 
of any of the three appropriate parameters -
magnetic field $B$, Doppler factor $\delta_{\rm j}$, and the maximum energy 
of accelerated electrons $E_{\rm max}$. 
However as it was argued by Pian et al.\ (\cite{Pian:98}), 
the shift of the synchrotron peak 
during these specific observations could hardly be attributed to the 
variation of the Doppler factor and/or magnetic field, but is caused rather by
an increase (by a factor of 10 or so) of the 
maximum energy of accelerated electrons.
On these grounds we may expect a significant 
hardening of the TeV spectrum as well.
However, due to the Klein-Nishina cross-section, the increase of 
$E_{\rm max}$ in the IC spectrum is expected to be substantially less
pronounced. 

It should be emphasized,
that during the whole period of 1997 the source was in a 
``high'' state with a TeV flux $\geq 1 \, \rm Crab$. 
It will be of utmost interest to use the IACT system to
study the Mkn~501 spectrum in a really low state, 
characterized by a TeV flux well below one Crab unit.

The second interesting feature of the {\it flux-selected} spectra 
averaged over almost 6 months of observations (Figure \ref{highlowfig}) 
is their smooth shape with power-law photon
index of about $\alpha\,=\,-$2.25 ($\mathrm{d}N_\gamma\,/\,\mathrm{d}E\,\propto\,E^{\alpha}$)
at energies between 1 TeV and several TeV, 
{\it but} with a gradual steepening at higher energies. 
We would like to make a comment concerning the 
implications of the steepening of the spectrum for the estimates of the 
diffuse extragalactic background radiation (DEBRA). 
If one interprets the lack of a cutoff in the $\gamma$-ray spectra of
both Mrk~421 (Zweerink et al.\ \cite{Zv:97}) and Mrk~501 
(Aharonian et al.\ \cite{Ahar:97a}) up to 10 TeV
as an indication for the absence of absorption in the DEBRA, an
upper limit on the energy density of DEBRA, $u_\epsilon=\epsilon^2 n(\epsilon)
\simeq 10^{-3} \, \rm eV/cm^3$ at $\lambda \sim 10 \, \mu \rm m$
can be derived from the condition of the 
transparency of the intergalactic medium
for 10 TeV $\gamma$-rays (Weekes et al.\ \cite{GRO4rev:97}). 
The recent studies of the problem, based on different 
assumptions about the spectral shape of the DEBRA,
give similar results (Stanev \&Franceschini \cite{Stanev:97};
Funk et al.\ \cite{Magn:97}; Biller et al. \cite{Bill:98};
Stecker \& De Jager \cite{Stec:98}). However, as it was
emphasized by Weekes et al.\ (\cite{GRO4rev:97}), 
the lack of an apparent cutoff in $\gamma$-ray
spectra  does not automatically imply negligible intergalactic absorption. 
Indeed, some infrared background models, like the {\it cold+hot dark matter} 
model of Macminn \& Primack (\cite{Pri:96}), predict a {\it modulation} rather than 
{\it cutoff} in the spectra of Mrk~421 and Mrk~501. 
The absorption results in a steeper observed spectrum, 
but even a power-law form could be conserved, at least up to 10~TeV.

The general tendency of gradual steepening of the spectra 
of Mrk~501 obtained in this paper could be formally interpreted as a result
of absorption in the intergalactic background radiation, which would allow 
to estimate the density of the DEBRA. Obviously this number could not be 
far from the above upper limit estimate. 
However, care should be taken in the interpretation of these results,
since the intrinsic spectra of the source are not properly understood
and probably several effects combine to steepen 
the TeV spectra of BL Lac objects.
\subsection{Temporal characteristics}
Our observations revealed flux variability on time scales 
$\Delta t_{\rm obs}$ of between 5 and 15\,h.
Due to causality and light travel time arguments the size of the 
$\gamma$-ray production region cannot exceed 
\begin{equation}
\label{tobs}
R =\Delta t_{\rm obs} c \, \delta_{\rm j} 
\simeq 3 \cdot 10^{15} \,\Delta t_{10}\, \delta_{\rm j} \, \rm cm, 
\end{equation}
with $\Delta t_{10} \,=\,\Delta t_{\rm obs}$/10\,h and
where $\delta_{\rm j}$ is the Doppler factor of the jet.
The condition that the source is optically thin with respect 
to photon-photon pair production, $\tau_{\gamma \gamma}  \le 1$,
results in  a {\it lower limit} on the Doppler factor of the jet 
$\delta_{\rm j}$, assuming that
the low-frequency photons are produced co-spatially in the quasi-isotropically
emitting cloud (blob): 
$\delta_{\rm j, min} \simeq 6 F_{-10}\-(\epsilon)^{1/6} \,\- \Delta t_{10}^{-1/6}$
(see e.g. Celotti et al.\ \cite{Cel:97}),
where $F_{-10}(\epsilon)=F_{\rm obs}(\epsilon)/$ $10^{-10} \, \rm erg/cm^2 s$ 
is the observed energy flux of the optical and the infrared 
photons at the observed energy $\epsilon=2 m_{\rm e}^2 \, c^4 \, \delta_{\rm j, min}^2/E_\gamma
\simeq 5 \, (\delta_{\rm j}/10)^2 \, (E_\gamma/10 \, \rm TeV)^{-1} \, \rm eV$
with width $\Delta \epsilon \simeq \epsilon$;
$E_\gamma$ is the energy of detected $\gamma$-ray photon. 
The characteristic fluxes of the optical and the infrared radiation from Mkn~501 of about 
$0.5 \cdot 10^{-10} \, \rm erg/cm^2 s$ (see e.g. Pian et al.\ \cite{Pian:98}), and 
the time variability of the 1-10 TeV $\gamma$-rays within 5 to 15~h 
obtained above, require a minimum Doppler factor in the order of 5.
Due to the weak dependence of $\delta_{\rm j, min}$
on $F_{\rm obs}$ and $\Delta t_{\rm obs}$, 
we can not expect a further significant strengthening of this 
lower limit on $\delta_{\rm j}$.

On the other hand, if the TeV $\gamma$-rays are produced by relativistic 
electrons which up-scatter their low-frequency synchrotron radiation 
(the so-called 
Syn\-chro\-tron Self \linebreak[4] Comp\-ton (SSC) scenario, see e.g. 
Ghisellini et al.\ \cite{GhisMD:1996}; Bloom \& Marscher \cite{BlMar:96};
Inoue \& Takahara \cite{IT:1996};
Mastichiadis \& Kirk \cite{Mast:97};
Bednarek \&  Protheroe \cite{BedPro:1997}),
the observed time variability of TeV $\gamma$-rays sets also a strong
{\it upper limit} on the Doppler factor, 
if one requires that the synchrotron and Compton
cooling time of the electrons is smaller than the flux variability time.
Indeed, the energy density of the low-frequency target 
photons in this model is estimated as 
$w_{\rm r} \simeq (d/R)^2 F_{\rm obs} c^{-1} \delta_{\rm j}^{-4}$, where
$d$ is the distance to the source ($\simeq 170 \, \rm Mpc$ 
for $H_0=60 \, \rm km/s\,Mpc$), $R$ is the size of the $\gamma$-ray 
production region  which is limited by Equation \ref{tobs},
but most probably cannot be significantly 
less than $R_{15}=R/10^{15} \, \rm cm$. 
Assuming that the synchrotron and Compton cooling time of electrons 
$t\,=\,[(4/3) \sigma_{\rm T} (w_{\rm r}+B^2/8\pi) E_{\rm e}/m_{\rm e}^2 c^4]^{-1} 
\simeq 15\, ((w_{\rm r}+B^2/8\pi)/1 \, \rm erg/cm^3)^{-1} \,$
$(E_{\rm e}/1 \,$$\rm TeV)^{-1} \, \, s$, where $B$ is the magnetic field in the jet,
does not exceed the flux variability time (in the frame of the jet)   
$\Delta t^{\prime} =\Delta t_{\rm obs} \, \delta_{\rm j}$,
we find 
\begin{equation}
\delta_{\rm j,max} \simeq 75\, F_{10}^{1/3}\,  
R_{\rm min,15}^{-2/3}\, \Delta t_{10}^{1/3}
\left(\frac{E_{\rm e}}{1\,{\rm TeV}}\right)^{1/3} \!\!
\left(\frac{F_{\rm X}}{F_{\rm TeV}}\right)^{1/3}\!\!
\end{equation}
where 
$F_{\rm X}/F_{\rm TeV}$ is the ratio of the energy flux emitted in the 
X-ray band and in the TeV band.
For characteristic values of
$F_{10} \simeq 0.5$, $R_{\rm min,15} \simeq 3$,
$\Delta t_{10} \simeq 1$, $E_{\rm e} \simeq 1$~TeV,
and $F_{\rm X}/F_{\rm TeV} \, \simeq \, 5$
(Pian et al.\ \cite{Pian:98})
one obtains $\delta_{\rm j, max} \simeq 50$.

In their different modifications, the inverse Compton (IC) models of
TeV radiation of BL Lac objects predict the correlation 
between the X-ray- and TeV-regimes which 
is indicated in Figures \ref{2dcorrfig} and \ref{corrfig}.
Albeit a correlation X/TeV is a strong argument
in favor of the common {\it electronic} origin of the parent particles
which produce synchrotron X-rays and IC $\gamma$-rays,
the fact of the correlation alone does not decide definitively 
between the electronic and hadronic nature of the primary
(accelerated) particles. For example in  
{\it Proton Blazar} type models (Mann\-heim \cite{Mann:93}), the bulk of
the nonthermal emission is produced at later stages of 
the proton-induced-cascade through the  same synchrotron and IC     
radiation of the secondary (cascade) electrons.

In fact, the short time variability of the keV/TeV-radiation
strongly argues in favor of electronic models.
Whereas the fast radiative 
(synchrotron and IC) cooling time of the electrons in the jet 
readily match the observed time-variability on a time scale of some hours,
the inelastic hadron interactions, both with ambient gas or photon fields
are rather slow processes and only become effective at 
very high target gas densities and/or photon densities, exceeding significantly
the typical values characterizing the $\gamma$-ray 
emitting jets in BL Lac objects (Schlickeiser \cite{Schl:96}; Sikora \cite{Sikora:97}). 
Nevertheless, presently the hadronic models cannot be ruled out unambiguously on the basis of
arguments concerning the time variability of the TeV-flux.
The rapid variability can be explained by geometrical effects, e.g., by 
anisotropies in the comoving frame of the jet (Salvati et al.\ \cite{Sal:98}), or in models
where the flares occur due to fast moving targets (gas clouds) which 
cross the beam of relativistic particles (Dar \& Laor \cite{Dar:97}). 
\begin{acknowledgements}
We thank the Instituto de Astrof\'{\i}sica de Canarias (IAC) for
supplying excellent working conditions at La Palma.
HEGRA is supported by the BMBF (Germany) and CYCIT (Spain).
The correlation analysis of the TeV flux and the RXTE ASM X-ray flux
has used data obtained through the High Energy Astrophysics
Science Archive Research Center Online Service,
provided by the NASA/Goddard Space Flight Center.
\end{acknowledgements}


\begin{thebibliography}{}
\bibitem[1997a]{Ahar:97a}{Aharonian, F.A., et al., 1997a, ApJ 327, L5}
%
\bibitem[1997b]{Ahar:97b}{Aharonian, F.A., et al., 1997b, Astropart. Phys.\ V.\ 6, I.\ 3-4, 343}
%
\bibitem[1997c]{Ahar:97c}{Aharonian, F.A., et al., 1997c. 
In: Proc. 4th Compton Symposium, AIP Conf. Proc.,
eds. C.Dermer, M.Strickman, and J.Kurfess, 1397}
\bibitem[1998a]{Ahar:98a}{Aharonian, F., et al., 1998a, submitted to Astropart. Phys.}
%
\bibitem[1998b]{Ahar:98b}{Aharonian, F., et al., 1998b, in preparation}
\bibitem[1997]{Barr:97}{Barrau, A., et al., 1997, astro-ph/9710259}
\bibitem[1997]{BedPro:1997}
{Bednarek, W., and  Protheroe, R. J., 1997, MNRAS 287, L9}
\bibitem[1997]{Bhat:97}{Bhat, C.L., et al., 1997. In: Procs. 
Towards a Major Atmospheric Cherenkov Detector~V, Kruger Park, South Africa}
\bibitem[1998]{Bill:98}{Biller et al., 1998, Phys. Rev. Lett. \,80, 2992-2995}
\bibitem[1996]{BlMar:96}{Bloom, S.D., and Marscher, A.P., 1996, ApJ 461, 657}
\bibitem[1997]{Brad:97}{Bradbury, S.M., et al., 1997, A\&A 320, L5}
\bibitem[1996]{Buck:96}{Buckley et al., 1996, ApJ 472, L9}
\bibitem[1998]{Buli:98}{Bulian, N., et al., 1998, Astropart.\ Phys.\ 8, 4}
\bibitem[1997]{Cata:97}{Catanese, M., et al., 1997, ApJ 487, L143}
\bibitem[1998]{Catan:98}{Catanese, M., et al., 1998, ApJ in press, astro-ph/9712325}
\bibitem[1997] {Cel:97} {Celotti, A., Fabian, A.C., and Rees, M.J., 1997, 
to appear in MNRAS, astro-ph/9707131}
\bibitem[1987]{Coll:87}{Collura, A., and Rosner, R., 1987, ApJ 315, 340}
\bibitem[1997]{Dar:97}{Dar, A., and Laor, A., 1997, ApJ 478, L5}
\bibitem[1997]{Daum:97}{Daum, A., et al., 1997, Astropart.\ Phys.\ 8, 1} 
\bibitem[1988]{Edel:88}{Edelson, R.A., Krolik, J.H., 1988, ApJ 333, 646}
\bibitem[1996]{Feg:96}{Fegan, D.J., 1996, Space Sci.\ Rev.\ 75, 137}
\bibitem[1997]{Fras:97}{Fra{\ss}, A., et al., 1997, Astropart.\ Phys.\ 8, 91}
\bibitem[1998]{Magn:97}{Funk et al., 1998,  Astropart.\ Phys.\ in press,
astro-ph/9802308}
\bibitem[1996]{Gaid:96}{Gaidos, J.A., et al., 1996, Nature 383, 319}
\bibitem[1996]
{GhisMD:1996}{Ghisellini, G., Maraschi, L., and Dondi, L., 1996, A\&AS 120, 503}
\bibitem[1998]{Haya:98}
{Hayashida, N., et al., 1998, astro-ph/9804043}
\bibitem[1998]{Hemb:98}
{Hemberger, M., 1998, PhD thesis, Universit\"at Heidelberg, Germany}
\bibitem[1985]{Hill:85}{Hillas, A.M., 1985. In: Proc. 19th ICRC, La
Jolla 3, 445}
\bibitem[1996]{IT:1996}{Inoue, S., and Takahara, F., 1996, ApJ 463, 555} 
\bibitem[1995]{Kono:95}{Konopelko, A., et al., 1995. In: 
Proc. Towards a Major Atmospheric Cherenkov Detector-IV,
ed. M.Cresti, 373}
\bibitem[1996]{konopelko96} Konopelko, A., et al., 1996, Astropart. Phys.\ 4, 199
\bibitem[1998]{Kono:98}{Konopelko, A., et al., 1998, submitted to Astropart. Phys.}
\bibitem[1993]{Mann:93}{Mannheim, K., 1993, A\&A 269, 67}
\bibitem[1997]{Mast:97}{Mastichiadis, A., Kirk, J.G., 1997, A\&A 320, 19}
\bibitem[1994]{Mirz:94}{Mirzoyan, R., et al., 1994, NIM A 315, 513}
\bibitem[1998]{Pian:98}{Pian, E., et al., 1998, ApJ 492, L17}
\bibitem[1996]{Petr:96}{Petry, D., et al., 1996, A\&A 311, L13}
\bibitem[1997]{Pueh:97}{P\"uhlhofer, G., et al., 1997, Astropart. Phys.\ 8, 101}
\bibitem[1992]{Punc:92}{Punch, M., et al., 1992, Nature 358, 477}
\bibitem[1996]{Pri:96}{Macminn, D., and Primack, J.R., 1996, Space Sci.\ Rev.\ 75, 413}
\bibitem[1997]{Prot:97}{Protheroe, R.J., et al., 1997, 25th ICRC,
Durban 1997, astro-ph/9710118}
\bibitem[1996]{Quin:96}{Quinn, J., et al., 1996, ApJ 456, L83}
\bibitem[1997]{Remi:97}{Remillard, R.A.\ and Levine, M.L., 1997, astro-ph/9707338}
\bibitem[1998]{Samu:98}{Samuelson, F.W., 1998, ApJ 501, L17}
\bibitem[1998]{Sal:98}{Salvati, M., Spada, M., Pacini, F., 1998, to appear in ApJ, 
astro-ph/9801049}
\bibitem[1996]{Schl:96}{Schlickeiser, R., 1996, Space Sci.\ Rev.\ 75, 299}
\bibitem[1997]{Sikora:97}{Sikora, M., 1997. In: 
Proc. 4th Compton Symposium, AIP Conf. Proc.,
eds. C.Dermer, M.Strickman, and J.Kurfess, 494}
\bibitem[1996]{ASCA_421:96}{Takahashi, T., et al., 1996, ApJ 470, L89}
\bibitem[1995]{Raut:95} Rauterberg, G., et al., 1995. In:
Proc. 24th International Cosmic Ray Conference, Rome, 3, 460
\bibitem[1998]{Stec:98}{Stecker, F.W. and  De Jager, O.C., 1998, A\&A in press}
\bibitem[1997]{Stanev:97}{Stanev, T., and Franceschini, A., 1998, ApJ 494, L159}
\bibitem[1997]{UUM:1997}{Ulrich, M.H., Maraschi, L., and Urry, C.M., 1997, ARA\&A 35, 445}
\bibitem[1995]{PadUry:95}{Urry, C.M.\ and Padovani, P., 1995, PASP 107, 803}
\bibitem[1997]{GRO4rev:97}{Weekes, T.C., et al., 1997. 
In: Proc. 4th Compton Symposium, AIP Conf. Proc.,
eds. C.Dermer, M.Strickman, and J.Kurfess, 361}
\bibitem[1998]{Wieb:98}{Wiebel, B., et al., 1998, A\&A 330, 389}
\bibitem[1997]{Zv:97}{Zweerink, J.A., et al., 1997, ApJ 490, L144}
\end{thebibliography}
\end{document}